# Learning Production Process Heterogeneity Across Industries: Implications of Deep Learning for Corporate M&A Decisions [†]

**Jongsub Lee**
Seoul National University (SNU) Business School

and

**Hayong Yun**
Eli Broad College of Business, Michigan State University

**Abstract**

Using deep learning techniques, we introduce a novel measure for production process heterogeneity across industries. For each pair of industries during 1990-2021, we estimate the functional distance between two industries' production processes via deep neural network. Our estimates uncover the underlying factors and weights reflected in the multi-stage production decision tree in each industry. We find that the greater the functional distance between two industries' production processes, the lower are the number of M&As, deal completion rates, announcement returns, and post-M&A survival likelihood. Our results highlight the importance of structural heterogeneity in production technology to firms' business integration decisions. (*JEL* G3, G34, L2, M1)

[†] Send correspondence to jongsub.lee@snu.ac.kr or yunhayon@msu.edu. The authors greatly appreciate helpful comments from discussions with Gordon Phillips. We thank Jung Hoon Han, and seminar and conference participants at Seoul National University (SNU) Business School and Michigan State University (Eli Broad College) for valuable comments, and Gayoung Koo for data assistance. Lee gratefully acknowledges financial support from the Institute of Finance and Banking and the Institute of Management Research at Seoul National University.

Industry characteristics are important for understanding the boundaries of a firm and the market for mergers and acquisitions (M&As). Industry characteristics could affect synergy value for an acquiring company (acquiror hereafter) when a firm expands its business boundary to the realm of the target company (target hereafter). Devos, Kadapakkam, and Krishnamurthy (2008) estimate synergy to be 10.03% of the combined company's equity value, of which operating synergy (e.g., cost reduction and margin improvement through economies of scale and greater pricing power, among many others) accounts for 8.38%, while tax savings explain the remaining 1.64%. Operating synergies tend to be higher in focused mergers, while tax savings comprise a large fraction of the gains in diversifying deals. These results are intuitive and economically meaningful as the cost of integration is likely to rise and negatively affects the value of synergy when firms with distinct production technologies and organizational capitals are combined under the one roof. However, it is unclear how to quantify the degree of heterogeneity in two firms' production decision processes, including their technological distances, cultural dissimilarity, and organization capital heterogeneity, because they are all latent by nature. This production process heterogeneity is important to identifying the cost of integration and resulting synergy value in diversifying mergers. Several studies attempt to capture the fundamental differences in asset composition and the underlying production technology using the text-based product characterization of acquiror and target (Hoberg and Phillips, 2010). We add to this important discussion by quantifying the underlying production process heterogeneity using novel deep learning techniques.

Which pairs of industries are more likely to be successfully integrated with greater synergy yet lower integration cost? The answer to this question requires a non-trivial extension of the simple dichotomous classifications of mergers in the literature (i.e., focused versus diversified M&As). To this end, we estimate the multi-stage production decision tree in each industry via deep neural network models. Specifically, we learn how neoclassical production inputs, such as total assets, capital expenditure, financial leverage, labor, asset tangibility, intangible assets, and free cash flow (let us denote this vector of production inputs in industry A as $x_A$) are mapped to the valuation output (Tobin's Q; let us denote this output in industry A as $y_A$).

These latent deep neural networks are illustrated in Figure 1. We estimate the weights of the $N \times 1$ input vector of a firm in industry A as $x_A = (x_1^{(A)}, x_2^{(A)}, ..., x_N^{(A)})$ for all layers of the production decision tree: 0,1,2,…, L. Optimal weights maximize the likelihood of the firm's valuation output in industry A: $y_A$. We estimate these optimal weights for all the layers of the production decision tree in each industry.

**Figure 1. A Simple Neural Network**

This figure graphically illustrates the industry-level production decision process as learned by a neutral network. As an example, we use industry A. The production inputs are denoted by a vector, $x_A = (x_1^{(A)}, x_2^{(A)}, \ldots, x_N^{(A)})$, and they are mapped to the valuation output $y_A$. The production decision tree has multiple layers: 0, 1, 2,..., L. Weights associated with the inputs in each layer are denoted by vectors: $W^{(0)}, W^{(1)}, W^{(2)}, \ldots, W^{(L)}$.

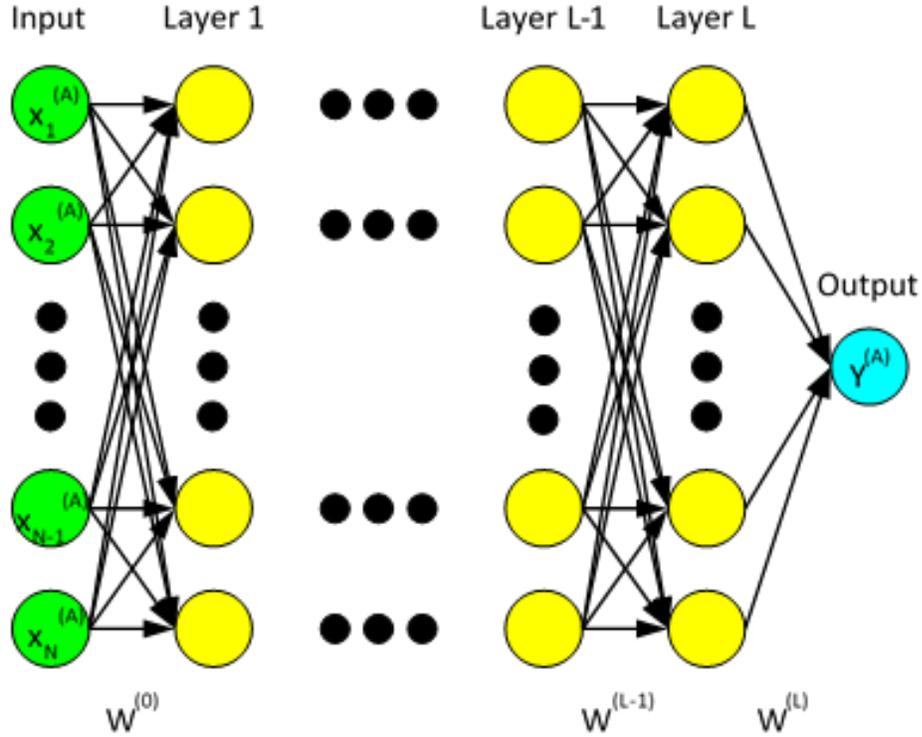

We then compute the distance in production decision processes between any pair of industries. For example, we first estimate the weights in industry A's production function, and force the weights from industry A to be the same for a second industry, B. We calculate the loss of fit in industry B's production process estimation under this weights constraint. Due to the forced weights, there should be a loss of fit in the constrained optimization in industry B. Hence, the mean squared errors from this constrained deep learning in industry B would be higher than the case in which we learn industry B's production process without such a weights constraint. This loss of fit can quantify the heterogeneity in two industries' underlying production processes. We estimate the loss of fit ratio with and without the weights constraint for all pairs of Fama-French 12 industries from 1990 to 2021. Using this novel concept of functional distances between industries, we quantify production process heterogeneity among acquirors and targets from various industries for the first time in the literature.

Following the merger synergy literature (Damodaran, 2005; Devos et al., 2009; Hoberg and Phillips, 2010; Deng, Kang, and Low, 2013), we hypothesize that the greater the functional distance between two industries' production profiles, the less synergetic value will be created in mergers between them. Business integration is more costly and less synergetic when two organizations with distinct production factors are combined into a single organization. For instance, it is difficult to create a large amount of synergy when a labor-intensive firm and a capital-intensive firm merge due to the high cost of business integration. Even when firms have the same production factors, those with different weights on the factors experience a more costly business integration than do firms with the same production factors and weights. For example, a capital-dependent firm with a positive weight on the capital factor would find it costly to be integrated with a firm with a negative weight on the same capital factor.

To further distinguish these two cases in diversifying mergers, i.e., (1) industries with different factors and weights and (2) industries with the same production factors but different weights, we introduce an additional deep learning algorithm known as Transfer Learning (TF learning). TF learning, contrary to our baseline deep learning algorithm, can conduct "layer extraction." In other words, TF learning can retain parts of layers from the trained industry (industry A) and apply the extracted layers to the second industry (B). We force all the layers in industry A except the last layer to be applied to industry B's production function estimation. Only the weights in the last layer of the decision tree in industry B are allowed to be freely determined. Given that most layers in the two industries' neural networks are configured and forced to be the same, the production factors that enter into the last layer, L, will be highly overlapped. What makes differences in their final production outcome will then be the weights on these common factors in the last layer of the decision tree.

In a simple two-factor multi-layer linear production process economy, we could theoretically demonstrate that firms with large functional distances in both Transfer Learning (TF Distance hereafter) and our baseline deep learning (Unadjusted Distance hereafter) have distinct factor structures and different weights. In sharp contrast, firms with a large Unadjusted Distance but a small TF Distance are associated with similar production factors yet different weights. The key intuition behind this theoretical prediction is that if production process heterogeneity can be explained by a simple reconfiguration of the weights in the last layer of the production decision tree, such firms should be grounded on similar factors. But the way those factors are employed by the firms in the last round of the production decision debate could significantly affect the firm's final valuation outcome.[1]

Based on this theoretical intuition, we further distinguish the two cases mentioned above in our data. For diversifying mergers, we posit that M&As between firms in industries with a large TF Distance tend to show a higher business integration cost than deals between industries with a large Unadjusted Distance. With more distinct factors and weights between acquiror and target, higher integration hurdles could exist, resulting in lower deal completion rates, poor stock market reactions upon the deal announcement, and a lower likelihood of survival after the two firms are combined into a single organization. We test all these predictions using the comprehensive U.S. M&A data compiled by SDC Platinum from 1990 to 2021.

We first examine whether industries with a large functional distance are associated with a lower number of M&As between them. In this industry-pair level analysis, we employ our baseline deep learning-based distance (Unadjusted Distance) as a main explanatory variable. We find strong support for our prediction. For a one standard deviation increase in log(Unadjusted Distance), there is a 30% reduction in log(Number of M&A Deals) from its sample standard deviation. When log(TF Distance) is used as an alternative explanatory variable, the economic significance is amplified to a 36% reduction in log(Number of M&A Deals). Both of our key right-hand-side (RHS) variables significantly explain our main dependent variable, log(Number of M&A Deals), at the 1% statistical significance level. The results hold not only in pooled panel regressions but also in year-by-year cross-sectional regressions. The results are largely consistent with our theoretical predictions.

Next, we dig further into deal-level analysis. We use and indicator of a deal successfully completed (i.e., a Deal Completion dummy) as our main left-hand-side (LHS) variable. When we control for various characteristics of M&As (Hostile, High Tech, Tender Offer, Stock Deal, Deal Size) and acquiror and target characteristics (Firm Size, Tobin's Q, Book Leverage, Cash Flow to Assets), we find robust and negative coefficients for both log(Unadjusted Distance) and log(TF Distance). These results are robust to year and industry fixed effects. All the results hold for both private and public acquirors/targets and only public acquirors/targets. Both of our distance measures explain a 7.1%- 7.7% reduction in the deal completion rate. They are statistically significant at the 1%- 5% level, depending on the specification of our probit model.

---

[1] In this regard, factors could be referred to as production technologies, whereas weights could be referred to as organizational structure/organizational capital. For example, see hierarchy versus polyarchy (Sah and Stiglitz, 1986), keeping authority vs. delegating authority (Dessein, 2002), and adaptive vs. inflexible organizational structures (Dessein and Santos, 2006).

Importantly, our functional distances could be highly correlated with Hoberg and Phillips' (2010, 2016) Text-based Network Industry Classifications (TNIC). TNIC identifies firms' end product markets based on the product descriptions in their 10-Ks. Using the TNIC3 scores provided in the Hoberg-Phillips data library for the 1990-2019 time period, we check the contemporaneous correlation between TNIC3 scores and our functional distances as well as their lead-lag relationship. We find that both of our functional distances are negatively correlated with TNIC3 scores at the 7% to 10% level, which implies that industries with similar end products (high TNIC3 scores) tend to have similar underlying production processes (low functional distances). The correlations are statistically significant at the 10% level. From the predictive regressions with one year lead and lag of these variables, we find that our functional distances predict TNIC3 scores, while TNIC3 scores also predict our functional distances. Both measures appear complementary and could jointly capture the dimensions that are relevant to merger synergy. When we run horse race regressions for log(Number of M&A Deals) using both measures (our functional distances and TNIC3 scores), we confirm that all of the variables significantly explain the log(Number of M&A Deals). Their effects are statistically significant at the 1% level. Their impacts are also economically meaningful (a 13% increase in merger intensity for a one standard deviation increase in TNIC3 score and a 35% reduction in merger intensity for log(TF Distance), respectively).

We move on to test our corollaries. By carefully comparing the effects of Unadjusted Distance and TF Distance, respectively, we test whether industries with different factors and weights (High Unadjusted Distance and High TF Distance industries) are much less likely to be integrated via M&As than industries with the same factor structure but different weights (High Unadjusted Distance but Low TF Distance industries). We expect a significantly negative log(Unadjusted Distance) effect, which becomes even more negative when the distance is interacted with log(TF Distance). We find one to three times amplified effects of log(Unadjusted Distance) when it is interacted with log(TF Distance). The results are consistent with our theoretical predictions from the linear stylized model. Our empirical evidence, together with the theoretical predictions, jointly provide a novel insight into the effect of production process heterogeneity on cross-industry merger activities – if acquiror and target are in industries with distinct factors (technologies) and weights (organizational capital), they have a higher cost of business integration than those in industries with common production factors yet different weights.

We further test M&A announcement returns and long-term survival rates for the combined organizations post-M&A using our distance measures. Using similar specifications to those employed by Deng, Kang, and Low (2013), we find for a one standard deviation increase in log(Unadjusted Distance) and log(TF Distance) results in -3% to -5% poorer stock market reactions over the (-1,+1) day window around the deal announcement date. Both effects are significant at the 1%-5% statistical significance level. For the long-term synergy effect, we test the survival rate of the combined organization over the one to two years following the deal completion date. For a marginal increase in both of our functional distance measures, there is an 8.7% to 13.7% reduction in the likelihood of survival. Overall, heterogeneous production decision processes between acquirors and targets are strongly correlated with poor valuation outcomes in both the short and long terms, indicating higher integration cost and lower synergy in mergers across heterogeneous industries in their production processes.

We conduct several robustness tests on the argument that alternative outcome variables could be more suitable for our deep learning models. We use operating profitability measured by the return on assets (ROA) as an alternative production outcome in our deep neural network models and re-run our analyses using the new set of production processes that we estimate. Our main results are virtually unchanged with this alternative production output specification.

We make several important contributions to the literature. To the best of our knowledge, we are the first to propose a way to quantify production process heterogeneity across industries using deep learning techniques. Production processes can contain not only the technological aspects of a firm but also encompass corporate culture, organizational capital, and other stakeholder-related characteristics (Sah and Stiglitz, 1986; Dessein, 2002; Dessein and Santos, 2006; Deng, Kang, and Low, 2013), all of which are latent by nature, and therefore difficult to estimate. Deep neural network models (which do not require ex-ante parametric assumptions) are ideal to uncover such latent and highly non-linear decision-making processes within a firm.[2]

---

[2] While the ordinary least squares (OLS) model can also fit nonlinearity, it requires a parametric assumption on the type of non-linear factor to be included in the model in advance (e.g., a factor, the factor-squared, and the factor-cubed, etc.). That is an econometrician's choice. However, in a deep neural network, the machine will find the most effective parametric functional form of non-linear factors without an econometrician's ex ante parametric assumption.

We not only estimate production decision trees using the latest machine learning techniques but also provide economic intuitions behind the estimated outcomes. We theoretically demonstrate that functional distances tend to be highest for industries with distinct production factors and weights, followed by industries with common production factors yet different weights. We also show that industries with similar production factors and weights have the lowest integration cost, and therefore tend to be merged in the market for M&As. Our strong empirical evidence together with our theoretical predictions add novel insights to the literature on the boundaries of the firm and merger theory (Jovanovic and Rousseau, 2001, 2002; Rhodes-Kropf and Robinson, 2008).

Jovanovic and Rousseau (2001, 2002) propose the Q-theory of mergers, which predicts that high Q industries acquire low Q industries due to valuation advantage. Rhodes-Kropf and Robinson (2008) propose an alternative theory, like-buys-like, in which firms in similar valuation cohorts are more likely to be integrated due to asset complementarity. Another stream of studies in the literature focuses on the industry merger wave and cross-industry merger dynamics (Harford, 2005; Hoberg and Phillips, 2010, 2016; Hoberg, Phillips, and Prabhala, 2014; Ahern, 2012; Ahern and Harford, 2014). All these studies jointly emphasize the importance of industrial organization dynamics and product market characteristics and their networks as main drivers of the observed patterns in the market for M&As. Using a novel identification strategy, we quantify production process heterogeneity across industries and show that it is an important determinant of cross-industry M&A activities.

Our findings also extend the simple dichotomous merger classifications in the literature (i.e., focused versus diversifying mergers). We show a significant variation in synergy value even among diversifying mergers. We find that the value of synergy critically varies with the underlying production technologies and organizational capital. In this regard, we propose a novel proxy for merger synergy in diversifying mergers and extend the merger synergy literature (Devos, Kadapakkam, and Krishnamurthy, 2008; Hoberg and Phillips, 2010; Deng, Kang, and Low, 2013). Devos, Kadapakkam, and Krishnamurthy (2008) highlight the tax benefits of diversifying mergers, while we show that mergers between distinct industries could also exhibit meaningful operational synergy, depending on the overlap in the underlying production decision processes between industries. Our measures also complement the text-based identification of firms' end product markets pioneered by Hoberg and Phillips (2010). Our results confirm that our functional distances capture the relevant dimensions of the firms' product making processes.[3] Our work is also closely related to Deng, Kang, and Low (2013) who emphasize the role of organizational capital, such as corporate social responsibility (CSR), to explain merger synergy.

## 1. Production Process Heterogeneity Across Industries: Theory and Measurement

### 1.1 A Simple Model

We explore a simplified two-factor linear production function to illustrate the difference in production decision processes between acquiror and target. Using this stylized model, we demonstrate what could explain production process heterogeneity across industries.

We consider three industries with different production functions from which acquiror and target pairs are matched: 1) a baseline industry, 2) an industry with the same latent production factors as the baseline industry that differs from it in the way the factors affect the final valuation output (i.e., weights), and 3) an industry with distinct production factors from the baseline industry. Production process heterogeneity is captured by deep learning across these three stylized groups.

Through this example, we highlight that acquiror and target firms that share common (latent) production factors may have different Unadjusted Distance but will not have significant difference in TF Distance. In contrast, acquiror and target firms that differ in (latent) production factors will result in significant TF Distance. As we will see in the subsequent sections, TF Distance tends to be a strong predictor of the likelihood of M&As because a large TF Distance implies a substantial difference in production processes between acquirors and targets. Such a difference cannot be easily accommodated without significantly compromising acquirors' existing organizational rules, decision-making processes, and corporate culture (Sah and Stiglitz, 1986; Dessein, 2002; Dessein and Santos, 2006).

***Model Setup***

Consider an economy with firms that have two-factor, three-layer hierarchy linear production functions as shown in Figure 2.

---

[3] Hoberg and Phillips (2016 JPE) mention that SIC codes are created based on production processes. The approach reflected in the SIC codes would also be complemented by our machine-based classifications on underlying production processes.

# Figure 2. Two-factor Linear Production Functions

This figure shows the two-factor, three-layer hierarchy linear production function example. These firms take capital (K) and labor (L) as input, and linear combinations of these factors are fed into the middle layer: $h_1 = w_{K1}K + w_{L1}L$ and $h_2 = w_{K2}K + w_{L2}L$. In turn, linear combinations of these middle layer outputs are fed to the top layer to make the production output: $y(K,L) = w_{H1}h_1 + w_{H2}h_2$.

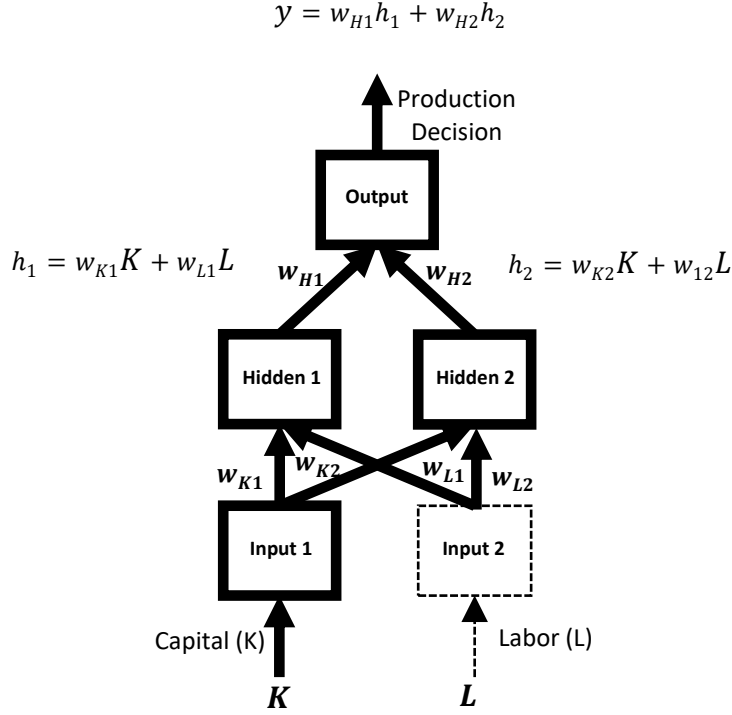

These firms take capital (K) and labor (L) as input. These factors are assumed to be uniformly distributed, i.e., $K \sim U(0,1)$ and $L \sim U(0,1)$. Linear combinations of these factors are fed into the middle layer (we refer $h_1$ and $h_2$ as latent factors):

$$h_1 = w_{K1}K + w_{L1}L$$

$$h_2 = w_{K2}K + w_{L2}L$$

In turn, linear combinations of these middle layer outputs are fed to the top to make the final production output:

$$y(K,L) = w_{H1}h_1 + w_{H2}h_2,$$

which can be explicitly expressed in terms of input factors of capital and labor as

$$y(K,L) = w_K K + w_L L,$$

where

$$w_K \equiv w_{H1}w_{K1} + w_{H2}w_{K2}$$

$$w_L \equiv w_{H1}w_{L1} + w_{H2}w_{L2}.$$

We consider three groups of firms (i.e., industries) based on the actual realization of outputs using input factors K and L. The first group is the baseline industry whose output depends on capital input and random noise:

$$y_1(K, L) = K + \varepsilon,$$

where

$$\varepsilon \sim \mathcal{N}(0, \sigma^2).$$

The second group's output also depends on capital input and random noise but the relationship between output and capital input differs from the baseline industry:

$$y_2(K, L) = -K + \varepsilon.$$

The second industry reflects the group of firms that has the same production factors as the baseline industry, but differs in how these factors are reflected in the production output.

The last group of firms has output that depends on a different factor from the baseline industry's, i.e., a labor factor, and random noise:

$$y_3(K, L) = L + \varepsilon.$$

The third industry represents firms that have different production factors and weights from those in the baseline industry.

### Industry Distance

In this subsection, we compute industry distances using two different estimation approaches: one without accommodation (from latent input factors to final output), referred to as Unadjusted Distance, and one with accommodation (from latent input factors to final output), referred to as TF Distance.

The objective of optimizing a business organization is to adjust the weights of each factor, $w \equiv \{w_{K1}, w_{K2}, w_{L1}, w_{L2}, w_{H1}, w_{H2}\}$, such that the expected mean squared errors between the model prediction, $y(K, L)$ and the actual observed outcome of the production, $y_i(K, L)$, are minimized

$$MSE_i = E\left[\int_0^1 \int_0^1 (y(K,L) - y_i(K,L))^2 dK\, dL\right].$$

It is straightforward to see that the mean squared error for the baseline industry is minimized when $w_1^* \equiv \{1,0,0,0,1,0\}$ with a resulting mean squared error of $\sigma^2$.[4]

A particularly interesting aspect arises in cross-industry M&As where acquirors and targets belong to different industries. It is straightforward to show that the mean squared errors of a firm in the baseline industry (group 1) and the other two industries (groups 2 and 3) are $\sigma^2, \frac{4}{3} + \sigma^2, \frac{13}{6} + \sigma^2$, respectively.[5] That is, the integration cost is minimal for within-industry mergers ($MSE_{NA,11} = \sigma^2$). The mid-range integration cost is between industries that share the same production factors, i.e., between groups 1 and 2 ($MSE_{NA,12} = \frac{4}{3} + \sigma^2$). Finally, the integration cost is highest between firms in industries with different input factors, i.e., groups 1 and 3 ($MSE_{NA,13} = \frac{13}{6} + \sigma^2$).

In practice, an acquiring firm can reasonably accommodate a target firm's production process when the underlying factors (latent factors in this model) significantly overlap. To reflect this flexibility in accommodating the target firm's production process, the second measure preserves organizational structure from the trained industry up to the middle of the hierarchy (i.e., fix $w_{K1}, w_{K2}, w_{L1}, w_{L2}$). But it allows accommodation at the top of the decision tree in the second industry (i.e., adjustment of $w_{H1}, w_{H2}$). It is clearly seen that the mean squared errors of a firm in the baseline industry and the

---

[4] To see this,
$$MSE_1 = E\left[\int_0^1 \int_0^1 (y(K,L;w_1^*) - y_1(K,L))^2 dK\, dL\right] = E\left[\int_0^1 \int_0^1 (K - (K+\varepsilon))^2 dK\, dL\right] = E\left[\int_0^1 \int_0^1 \varepsilon^2 dK\, dL\right] = E[\varepsilon^2] = \sigma^2.$$
I.e., it cancels out the input factor, K, and is only left with contributions from random noise.

Note that this is not the only mean squared error minimizing solution. Other solutions are $w_1^{**} \equiv \{\eta, 1-\eta, 0,0,1,1\}$, $w_1^{***} \equiv \{1,1,0,0,\eta, 1-\eta\}$, $w_1^{****} \equiv \{\alpha, \beta, 0,0, \frac{1}{\alpha}, \frac{1}{\beta}\}$, where $0 \leq \eta \leq 1$, $0 < \alpha < 1$, $0 < \beta < 1$. In this example, we will focus on $w_1^*$ to simplify exposition.

[5] The MSE of the group 1 model applied to group 1 input-output data is
$$MSE_{NTF,11} = E\left[\int_0^1 \int_0^1 (K - (K+\varepsilon))^2 dK\, dL\right] = E\left[\int_0^1 \int_0^1 \varepsilon^2 dK\, dL\right] = E[\varepsilon^2] = \sigma^2.$$
The MSE of the group 1 model applied to group 2 input-output data is
$$MSE_{NTF,12} = E\left[\int_0^1 \int_0^1 (4K^2 - 4K\varepsilon + \varepsilon^2) dK\, dL\right] = E\left[\frac{4}{3} - 2\varepsilon + \varepsilon^2\right] = \frac{4}{3} + \sigma^2.$$
The MSE of the group 1 model applied to group 3 input-output data is
$$MSE_{NTF,13} = E\left[\int_0^1 \int_0^1 (K - (L+\varepsilon))^2 dK\, dL\right] = E\left[\int_0^1 \left(\frac{4}{3} + L^2 + \varepsilon^2 - L + \varepsilon - 2L\varepsilon\right) dL\right] = \frac{13}{6} + \sigma^2.$$

other two groups with this additional adjustable layer are $\sigma^2, \sigma^2$, and $\frac{4}{3} + \sigma^2$, respectively.[6] When layer adjustments are allowed, integration costs are generally lower than in a case without such accommodations. The resulting integration costs are the lowest under the layer adjustment for same industry mergers ($MSE_{TF,11} = \sigma^2$) and cross-industry mergers with the same production factors ($MSE_{TF,12} = \sigma^2$). The integration cost is highest between firms from industries with a completely different production factor structure, i.e., groups 1 and 3 ($MSE_{TF,13} = \frac{13}{6} + \sigma^2$).

The above model further highlights the sensitivity of business integration costs with and without layer adjustment. The integration cost without layer adjustment is more sensitive to the difference in input factors across industries (our baseline deep learning) than those with layer extraction (TF learning). The latter approach is, therefore, more likely to capture salient differences rather than subtle differences in the underlying production processes across industries.

## 1.2 Empirical Estimation of Production-Function-Based Industry Distances

We define a measure of industry distance that captures the cost of integration between two firms in a merger transaction.[7] Motivated by the stylized example in the previous section, we conjecture that firms that have very different production processes are likely to have different organizational structures, decision making processes, and cultures, which hinder smooth accommodation of the newly acquired target to the acquiror. If the difference in production processes is very large between acquiror and target, the cost of integration may outweigh the benefits, leading to negative operating synergy. This would also lead to a deterring M&A attempt and an adverse impact on the long-term success rate

---

[6] The MSE of the group 1 model applied to group 1 input-output data is transfer learning $w_1^*$ to fit $y_1(K,L) = K + \varepsilon$ using model $y(K,L; w_{TF,1} = w_1^*) = w_K K + w_L L = K$, which is the same as no transfer learning case because the weights are already optimal, and no transfer learning is needed, i.e., $MSE_{TF,11} = \sigma^2$.

The MSE of the group 1 model applied to group 2 input-output data is transfer learning $w_1^*$ to fit $y_2(K,L) = -K + \varepsilon$ using model $y(K,L) = w_K K + w_L L$. The optimal weights for $y_2(K,L)$ are $w_2^* \equiv \{1,0,0,0,-1,0\}$, which can be achieved by altering the last layers' weights of $w_1^* \equiv \{1,0,0,0,1,0\}$ to $w_{TF,2} = \{1,0,0,0,-1,0\}$, i.e., change to $w_{H1} = -1$. Hence, $y(K,L; w_{TF,2}) = w_K K + w_L L = -K$. As a result, $MSE_{TF,12} = E\left[\int_0^1 \int_0^1 (-K - (K + \varepsilon))^2 dK\, dL\right] = E[\varepsilon^2] = \sigma^2$.

The MSE of the group 1 model applied to group 3 input-output data is transfer learning $w_1^*$ to fit $y_3(K,L) = L + \varepsilon$ using model $y(K,L; w_{TF,3}) = w_K K + w_L L$. Since the true output, $y_3(K,L)$, only depends on labor and not capital, the second best (because the first layer weights from labor are zero and there is no way to incorporate labor input effects) way to minimize mean squared error loss is to suppress the capital input effects which only adds noise irrelevant to the true output factor (labor), $w_{TF,3} = \{1,0,0,0,0,0\}$, i.e., change to $w_{H1} = 0$, and we get $y(K,L; w_{TF,3}) = 0$.

$$MSE_{TF,13} = E\left[\int_0^1 \int_0^1 (0 - (L + \varepsilon))^2 dK\, dL\right] = \frac{4}{3} + \sigma^2.$$

[7] Hoberg-Phillips's TNIC-3 scores focus on product is suitable for identifying competitors and the benefits of mergers; our focus on production function is suitable for identifying the costs associated with merging two organizations.

of the M&A. For these reasons, we quantify the difference in production functions between every pair of industries, which we call functional distances between them.

Instead of a structural approach (e.g., the Cobb-Douglas function) to estimate acquirors' production functions in each industry, we use deep neural network learning that does not require any parametric assumptions on the functional form of the production technology. To represent a firm's production decision process (the relation between input factors and production outcome), we use acquirors' input ($x_A$)-output ($y_A$) data to fit the acquirors' production function, $f_A()$, in each year for all the Fama-French 12 industries using a deep neural network,

$$f_A\left(x_A; w_A^{(0)}, \cdots, w_A^{(L)}\right) = \hat{y}_A.$$

For production input vector $x_A$, we consider the natural logarithm of total assets, capital expenditures divided by total assets, short term debt divided by total assets, long term debt divided by total assets, employment divided by total assets, property plant and equipment divided by total assets, advertisement expenses divided by total assets, and R&D expenditure divided by total assets. For the production output, $y_A$, we use the natural logarithm of Tobin's Q. In addition to this valuation-based outcome, we later consider an alternative production outcome, ROA (operating income divided by total assets), and show the robustness of our main results.

To eliminate industry specific effects in each year, we consider a variable, $x_{ijt}$, for industry $i$ in firm $j$ in year $t$. We compute the industry average of this variable, $\bar{x}_{it} = \frac{1}{N_{it}} \sum_{k=1}^{N_{it}} x_{ijt}$, where $N_{it}$ is the number of firms in industry $i$ in year $t$. Using these industry average values, we adjust for our original variables:

$$\tilde{x}_{ijt} = x_{ijt} - \bar{x}_{it},$$

where $x_{ijt}$ is Tobin's Q, ROA, log of assets, short term debt/assets, long term debt/assets, PPENT/assets, employment/assets, advertisement expense/assets, and R&D expense/assets. That is, our input and output variables in the acquirors' production function estimation are deviations from the annual industry average values rather than their raw values.[8]

---

[8] Using raw values instead of the deviation from annual industry means gives quantitatively similar results, but we focus on the deviation from the annual mean since this controls for cross-industry shifts in levels and standardizes comparison across industries.

The fitting criteria is to find neural weights that minimize the mean squared errors between the neural network-based production function ($\hat{y}_A$) and the actual observed production outcome ($y_A$):

$$\min_{w_A^{(0)},\cdots,w_A^{(L)}} \mathcal{L}_A\left(x_A; w_A^{(0)}, \cdots, w_A^{(L)}\right) = \min_{w_A^{(0)},\cdots,w_A^{(L)}} MSE(\hat{y}_A) =$$

$$\min_{w_A^{(0)},\cdots,w_A^{(L)}} \sqrt{\frac{1}{N}\sum_{j=1}^{N}(y_{A,j} - \hat{y}_{A,j})^2}.$$

We retain the resulting neural weights, $w_A^{(0)}, \cdots, w_A^{(L)}$, which fit the production functions of the firms in each acquiror's industry. Once we estimate an acquiror's industry-level production function, we measure the target industry's goodness of fit by computing the mean squared errors of the target industry's production function, while we fix its neural weights to be the same as those we already estimated for the acquiror's industry, A:

$$MSE\left(y_T, x_T; w_A^{(0)}, \cdots, w_A^{(L)}\right) = \sqrt{\frac{1}{N}\sum_{j=1}^{N}(y_{T,j} - \hat{y}_{U,T,j}(x_T))^2},$$

where $y_{T,j}$ is the actual production outcome of firm $j$ in the target industry, and $\hat{y}_{U,T,j}$ is the model estimate of firm j's production function using the trained weights from the acquiror's industry ($w_A^{(0)}, \cdots, w_A^{(L)}$):

$$f_A\left(x_T; w_A^{(0)}, \cdots, w_A^{(L)}\right) = \hat{y}_{U,T}(x_T).$$

We then define Unadjusted Distance (see Figure 3) between the acquiror's and the target's industries as the normalized mean squared errors obtained above, $MSE\left(y_T, x_T; w_A^{(0)}, \cdots, w_A^{(L)}\right)$ by the mean squared errors of the target industry itself without the same weights constraint from the acquiror's industry, i.e., $MSE\left(y_T, x_T; w_T^{(0)}, \cdots, w_T^{(L)}\right)$:

$$d_U(y_A, y_T) = \frac{MSE\left(y_T, x_T; w_A^{(0)}, \cdots, w_A^{(L)}\right)}{MSE\left(y_T, x_T; w_T^{(0)}, \cdots, w_T^{(L)}\right)}.$$

This measure captures the degradation of estimation efficiency in the constrained optimization when we learn the production process of the target industry while imposing the organizational weights learned from the acquiror's industry. It should also be noted that these industry distances are asym-metric, reflecting the viewpoint of acquirors who apply their production functions to the target companies' production functions and see the potential cost of integration when they make M&A decisions (e.g., a complex firm acquiring a target with a simple task could have a different industry distance from a simple firm acquiring a complex target).

### Figure 3. The Network Structure for Unadjusted Industry Distance

This figure shows how Unadjusted Distance is estimated. The top panel shows training a production function for the acquiror's industry. We use the acquiror's industry's input-outcome data to fit its production function using a deep neural network, $f_A(x_A; w_A^{(0)}, \cdots, w_A^{(L)}) = \hat{y}_A$. The fitting criteria are to find neural weights that minimize the mean squared distance between the neural network represented production function and the actual production outcome:

$$\min_{w_A^{(0)}, \cdots, w_A^{(L)}} \mathcal{L}_A(x_A; w_A^{(0)}, \cdots, w_A^{(L)}) = \min_{w_A^{(0)}, \cdots, w_A^{(L)}} MSE(\hat{y}_A) = \min_{w_A^{(0)}, \cdots, w_A^{(L)}} \sqrt{\frac{1}{N}\sum_{j=1}^{N}(y_{A,j} - \hat{y}_{A,j})^2}.$$

We retain the resulting neural weights, $w_A^{(0)}, \cdots, w_A^{(L)}$, that fit the production function of firms in the acquiror's industry. The bottom panel shows how we obtain the mean square errors of production functions using the target's industry's data:

$$MSE(y_B, x_B; w_A^{(0)}, \cdots, w_A^{(L)}) = \sqrt{\frac{1}{N}\sum_{j=1}^{N}(y_{B,j} - \hat{y}_{U,B,j}(x_B))^2}$$

Where $y_{B,j}$ is the actual production outcome of firm $j$ in the target industry, and $\hat{y}_{U,B,j}$ is the model estimate of firm j's production function which uses trained weights optimized for firms in the acquiror industry:

$$(x_B; w_A^{(0)}, \cdots, w_A^{(L)}) = \hat{y}_{U,B}(x_B)$$

$$MSE(y_B, x_B; w_A^{(0)}, \cdots, w_A^{(L)}) = \sqrt{\frac{1}{N}\sum_{j=1}^{N}(y_{B,j} - \hat{y}_{U,B,j}(x_B))^2}$$

The unadjusted distance between the acquiror's industry and the target's industry is the ratio of mean squared errors obtained above, $MSE(y_B, x_B; w_A^{(0)}, \cdots, w_A^{(L)})$, with respect to the mean squared errors when the model is fitted for firms in the target industry, $MSE(y_B, x_B; w_B^{(0)}, \cdots, w_B^{(L)})$:

$$d_U(y_A, y_B) = \frac{MSE(y_B, x_B; w_A^{(0)}, \cdots, w_A^{(L)})}{MSE(y_B, x_B; w_B^{(0)}, \cdots, w_B^{(L)})}.$$

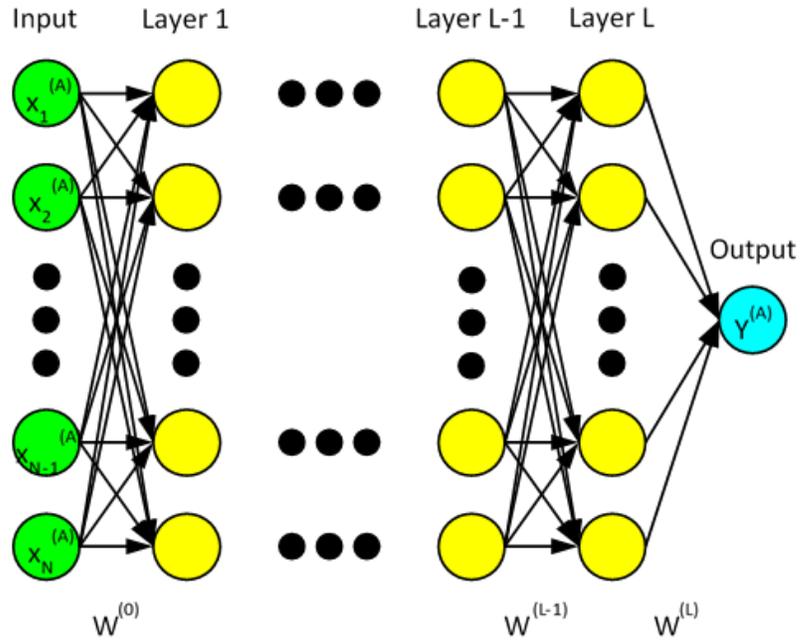

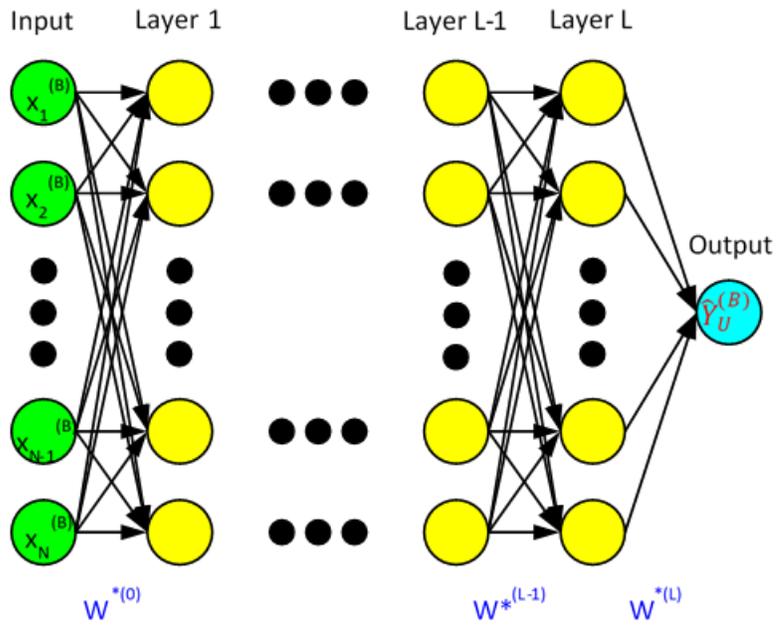

While Unadjusted Distance captures the differences between acquirors' and targets' production functions, it may be too sensitive, picking up even small differences that might be easily accommodated in practice. To capture only the significant differences in production functions between acquiror and target, we additionally consider Transfer Learning-Based (TF) Distance, which is estimated through a similar approach except the following condition: we use the trained industry's weights (i.e., the acquiror's production function weights) only up to the second to last layer of the deep neural network of the test industry (i.e., the target's industry). In other words, we keep $w_A^{(0)}, \cdots, w_A^{(L-1)}$, while we allow the weights in the last layer, $w_A^{(L)}$, to be freely adjustable. Under this relaxed weights constraint, we minimize the mean squared errors of the target industry's production function estimation (see Figure 4),

$$\min_{w_A^{(L)}} \mathcal{L}_B\left(x_B; w_A^{(0)}, \cdots, w_A^{(L-1)}\right) = \min_{w_A^{(L)}} \sqrt{\frac{1}{N}\sum_{j=1}^{N}\left(y_{B,j} - \hat{y}_{TF,B,j}\right)^2}$$

The resulting estimated production function is denoted as

$$f_A\left(x_T; w_A^{(0)}, \cdots, w_T^{(L)}\right) = \hat{y}_{TF,T}.$$

We normalize the resulting mean squared errors by the mean squared errors of the production function of the target industry itself without any weights restriction, $MSE\left(\hat{y}_T, y_T, x_T; w_T^{(0)}, \cdots, w_T^{(L)}\right)$. This leads to the following TF Distance:

$$d_{TF}(y_A, y_T) = \frac{MSE\left(\hat{y}_{TF,T}, y_T, x_T; w_A^{(0)}, \cdots, w_A^{(L-1)}, w_T^{(L)}\right)}{MSE\left(\hat{y}_T, y_T, x_T; w_T^{(0)}, \cdots, w_T^{(L)}\right)}.$$

### Figure 4. The Network Structure for Transfer Learning-Based Industry Distance

This figure shows how transfer learning-based industry distance is estimated. The top panel shows training a production function for the acquiror's industry, which is the same as in the case of Unadjusted Distance. The bottom panel shows obtaining the mean square errors of the production function using target industry data. In the TF Distance, we use the trained weights from Step 1 only up to the second to the last layer of the deep neural network, $w_A^{(0)}, \cdots, w_A^{(L-1)}$, and allow the weights of the last layer to adjust to minimize mean squared errors using Industry B production data

$\min_{w_A^{(L)}} \mathcal{L}_B\left(x_B; w_A^{(0)}, \cdots, w_A^{(L-1)}\right) = \min_{w_A^{(L)}} \sqrt{\frac{1}{N}\sum_{j=1}^{N}(y_{B,j} - \hat{y}_{TF,B,j})^2}.$
The resulting estimated production function is denoted as, $f_A(x_B; w_A^{(0)}, \cdots, w_B^{(L)}) = \hat{y}_{TF,B}.$
Using the resulting mean squared error with respect to $MSE(\hat{y}_B, y_B, x_B; w_B^{(0)}, \cdots, w_B^{(L)})$ gives the transfer learning-based industry distance

$$d_{TF}(y_A, y_B) = \frac{MSE(\hat{y}_{TF,B}, y_B, x_B; w_A^{(0)}, \cdots, w_A^{(L-1)}, w_B^{(L)})}{MSE(\hat{y}_B, y_B, x_B; w_B^{(0)}, \cdots, w_B^{(L)})}.$$

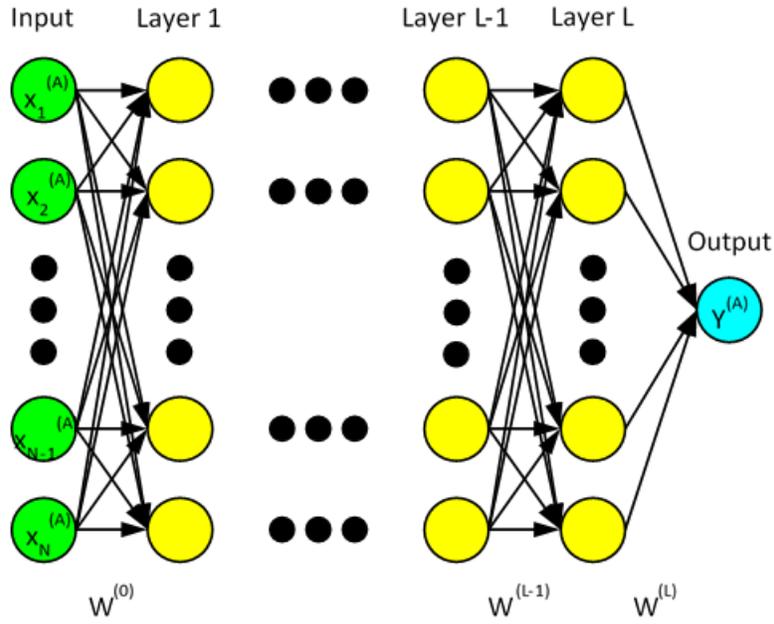

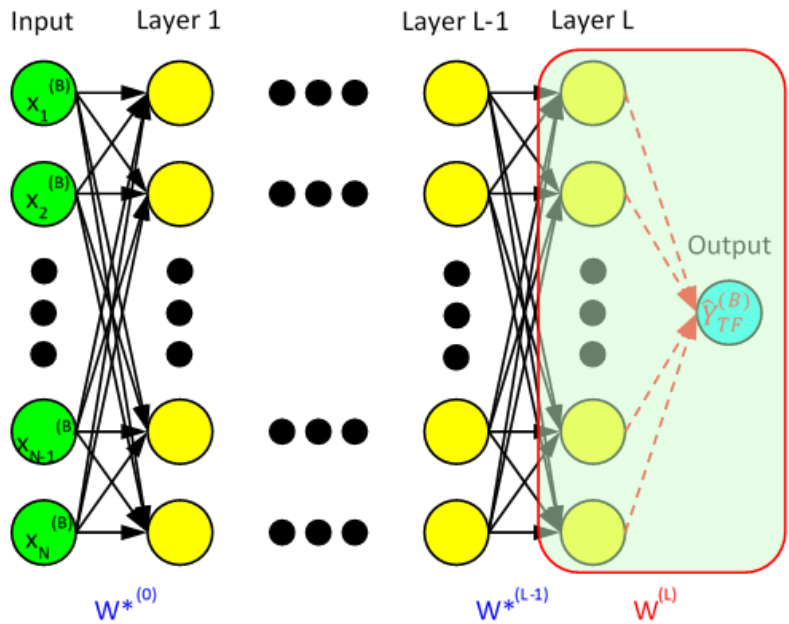

## 2. Hypotheses Development

Based on the stylized model and the identification strategy for industry-level functional distances, in the following sections we test our main propositions and corollaries. In our empirical analyses, we consider both Unadjusted Distance and TF Distance as our key explanatory variables and conduct various analyses that examine merger intensity, announcement effects, and post-merger long-term performance consequences.

**Proposition 1**: *For high Unadjusted Distance (or TF Distance), the outcome of the production process (i.e., firms' end products) could differ between two industries. TNIC3 scores are, therefore, expected to be negatively correlated with the functional distances between a pair of industries.*

The first proposition ensures that we are capturing the relevant dimension of the firms' production processes. Distinct production factors and weights are likely to be associated with the different product markets in which acquirors and targets operate.

Moreover, substantial differences in production decision processes are expected to be associated with a higher integration cost between the two organizations. This leads to the following prediction:

**Proposition 2**: *For high Unadjusted Distance (or TF Distance), firms in two industries are less likely to be integrated through M&As.*

We note that TF Distance captures more salient differences in the underlying production processes between two industries. As the reconfiguration of weights could accommodate two industries' production decision processes easily in certain cases, industries with a high TF Distance between them are more likely to be associated with distinct production factors and weights. This intuition leads to the following two corollaries:

**Corollary 1**: *If a pair of industries has a high Unadjusted Distance but a low TF Distance, their factor structure is more likely to overlap.*

**Corollary 2**: *If a pair of industries has both a high Unadjusted Distance and a high TF Distance, they are more likely to have a heterogeneous factor structure.*

Given all the predictions above, we expect the following empirical regularities in the market for M&As:

**Proposition 3**: *The negative effect of TF Distance on M&A likelihood will be larger than that of Unadjusted Distance.*

**Corollary 3**: *The negative effect of Unadjusted Distance on M&A intensity is magnified as TF Distance increases. The point estimate of the interaction term between Unadjusted Distance and TF Distance on M&A intensity is expected to be negative.*

Now we bring these propositions and corollaries to our data.

## 3. Data

### 3.1 Sample Construction

The main sample for this study consists of the annual panel of 144 Fama-French 12 industry pairs (Fama and French (1997) from 1990 to 2021.[9] There are a total of 4,608 observations. Industry distances are constructed using the financial information of the U.S. firms in the Compustat annual database. We require a firm to have total assets greater than $10 million. We also require no missing information for our production input and output variables, including total assets, Tobin's Q (total shares outstanding, year-end stock price, market value of common equity, deferred taxes), book leverage (debt in current liabilities, long term debt), ROA (operating income before depreciation), labor (number of employees), asset tangibility (property plant and equipment), intangible assets (advertising expenses, R&D expenses), and free cash flow (operating income before depreciation, total interest and related expenses, total income taxes, capital expenditures). This data requirement results in 286,469 firm-year observations from annual Compustat data for training and testing production functions. Using these data, we estimate industry distance measures for all 144 Fama-French 12 industries pairs in each year during our sample period. To inspect the relationship between our industry-level functional distances and Hoberg and Phillips's (2010, 2016) firm rival scores, we merge the average TNIC3 scores for each Fama-French 12 industries pair to our annual panel data. TNIC3 scores are available from 1990 to 2019 (4,300 observations). We obtain them from Hoberg and Phillips's data library (https://hobergphillips.tuck.dartmouth.edu/).

We also conduct M&A deal-level analyses using our industry distance measures. We obtain deal-level M&A information from the SDC platinum database and merge it with acquirors and targets' firm financial information from Compustat from 1990 to 2021. For this study, we require the deals to have the following information (SIC codes of acquiror and target, whether acquiror and target were in a high technology industry, whether the merger was hostile, whether it used tender offers, whether the deal was financed partly by stocks, deal size, and acquiror's equity value). The deals should have clear intentions (either friendly or hostile, which covers more than 90% of the data). These are all standard data filters used in the literature. This data filter leaves 35,613 deals from 1990 to 2021 (there are 267,168 deals with non-missing deal characteristics except deal size, which limits the final sample to 35,613 deals). We also consider deals that involve publicly traded acquirors and deals that involve publicly traded acquirors and publicly traded targets. Using SDC CUSIP, we match the firms (acquirors and targets) to the Compustat universe. We have 12,365 deals for the public acquirors subsample, and 3304 deals for the public acquirors and public targets subsample.

---

[9] Industry data are compiled from Kenneth E. French's data library (https://mba.tuck.dartmouth.edu/pages/faculty/ken.french/data_library.html).

## 3.2 Variable Description

The main variables of interest for this study are industry distance measures: Unadjusted Distance and TF Distance. For both measures, we use a deep neural network to find production functions; the network maps production input (deviation from the industry average of the natural logarithm of total assets, capital expenditures divided by total assets, short term debt divided by total assets, long term debt divided by total assets, employment divided by total assets, property plant and equipment divided by total assets, advertisement expenses divided by total assets, and R&D expenditure divided by total assets) to production output (deviation from the industry average of the natural logarithm of Tobin's Q for the baseline distance measure, and ROA for the alternative distance measure). The industry distance measures are asymmetric because, for example, applying an acquiror production function that deals with comprehensive tasks to a target's production function that requires simple tasks is easy, but not vice versa. We log-transform the two functional distances and denote them by log(Unadjusted Distance) and log(TF Distance), respectively.

In order to study M&A intensity between a pair of Fama-French 12 industries, we compute the Number of M&A Deals between each pair of Fama-French 12 industries reported in the SDC Platinum data. Log(Number of M&A Deals) is the natural logarithm of the variable. For the M&A deal-level analyses, we construct an indicator for M&A deal completion (Deal Completed dummy). As controls in our deal-level analyses, we consider the following deal and acquiror/target company characteristics widely used in the literature (Deng, Kang, and Low, 2013, among many others). Diversify is an indicator for a diversifying merger (the acquiror's two digit SIC code differs from that of the target's). Hostile is an indicator for a hostile merger, which takes a value of one if SDC Platinum records the deal as a hostile merger and zero otherwise. High Tech is an indicator that takes a value of one if both acquiror and target are in high technology sectors and zero otherwise (i.e., SDC Platinum records them as "Primary Business not Hi-Tech"). Tender Offer is an indicator for tender offer deals. Stock Deal is an indicator for deals that are (partly) financed by stock according to the report in SDC. Relative Deal Size is the deal value reported in SDC normalized by the acquiror's market capitalization.

Various acquirors' and targets' characteristics are compiled from Compustat. Firm size is the natural logarithm of total assets. Tobin's Q is the ratio of total assets plus market value of equity (total shares outstanding times year-end stock price) minus the total value of common equity minus deferred taxes to total assets. Book Leverage is debt in current liabilities plus total long-term debt divided by total assets. Cash Flow to Assets is operating income before depreciation minus total interest related expenses minus total income taxes minus capital expenditures divided by total assets.

TNIC3 scores (Hoberg and Phillips, 2010, 2016) are at the firm level, while our industry distances are measured at the Fama-French 12 industries pair level. Hence, we compute the average scores for each Fama-French 12 industries pair in each year.

### 3.3 Summary Statistics

Table 1 reports the summary statistics for all the variables used in this paper. Panel A shows the mean, standard deviation, and 5/50/95 percentile values of the Fama-French 12 industries pairs by year level characteristics. Both Unadjusted Distance and TF Distance have mean values greater than one (1.32 for Unadjusted Distance and 1.285 for TF Distance, respectively). This is expected and consistent with the Lagrange multiplier-type intuition of these variables. Theoretically, these distances should equal one if they measure the distance of an industry from itself. Empirically, these distances could fall below one as the MSE uses out of sample data. But most of the time in our sample, they are more than or equal to one; the 5th percentile of this variable is 1.0 for Unadjusted Distance and 1.074 for TF Distance, respectively. Both distances show significant variations. Their standard deviations are 27.8% (Unadjusted Distance) and 21.4% (TF Distance). Their upper bounds can also be very large, exceeding 200% as illustrated by their 95th percentiles: 178.5% for Unadjusted Distance and 171.5% for TF Distance. Note that the average and 95th percentile values of TF Distance are smaller than those of Unadjusted Distance because TF Distance tends to be large only for substantial differences in acquirors' and targets' production functions. It tends to stay close to one most of the time, whereas Unadjusted Distance can be large when there are even small differences in the production functions of acquirors' and targets' industries. For regressions, we use log-transformed distance measures to address potential skewness in these variables.

Our baseline specification uses log(Q) as the production output, which is common in q-theory. However, we also consider an alternative specification that uses ROA as the production output. This alternative specification based functional distances are summarized in Panel A as well. Both Unadjusted Distance and TF Distance have larger average values than the baseline log(Q) based specification (1.552 for Unadjusted Distance and 1.338 for TF Distance). They also have greater variations; the standard deviations are 58.1% for the ROA-based Unadjusted Distance and 30.7% for the ROA-based TF Distance.

On average, there are 65 deals in a year for a Fama-French 12 industries pair (see the mean value of Number of M&A Deals). But there is substantial variation with the standard deviation of the variable being 208 deals.

In Panel B, we summarize the variables at the M&A deal level. The Deal Completed dummy has an average of 90% for all three subsamples (89.2% for all deals, 91.7% for the public acquiror subsample, and 88.7% for the public acquiror-public target subsample) with significant variation – the standard deviations are 31%, 27.5%, and 31.7%, respectively.

A little over half of the deals are within two-digit SIC codes, but there is a substantial fraction of cross-industry mergers; the average of the Diversify dummy is 48.1% (all deals), 38.4% (public acquiror deals), and 34.6% (public acquiror-public target deals). The fraction of hostile mergers tends to be small (0.9% for all deals and public acquiror deals), but is somewhat larger for public acquiror-public target deals (2.6%). Many deals seem to occur in high technology sectors, with the average High Tech deal dummy being 21.4% (all deals), 34% (public acquiror deals), and 42.6% (public acquiror-public target deals). The public acquiror-public target deals tend to use tender offers more often (17.4%) than other types of deals (5% for all deals and 5.4% for public acquiror deals). The Stock Deal dummy is around 2% across all three samples. In public acquiror-public target deals, the acquiror Firm Size (mean of 7.529) tends to be larger than that of the target (mean of 5.499), but Tobin's Q tends to be larger for the target (mean of 4.774) than the acquiror (mean of 2.362). In public acquiror-public target deals, both acquiror and target have similar book leverage ratios, with an average around 23%. Their Cash Flow to Assets ratios are also similar, around -1.5% on average. These are consistent with the like-buys-like theory of mergers proposed by Rhodes-Kropf and Robinson (2008).

In Panel C, we summarize the Hoberg-Phillips TNIC3 Score. The variable has an annual average for each Fama-French 12 industries pair of 0.03 with a standard deviation of 2.1%. It varies from 1.2% to 5.5% in the 5th-95th percentile range.

In Panel D, we report the (-1,0) and (-1,+1) CARs for both equal- and value-weighted cases, where we find the equal-weighted CAR (t-1 to t+1 mean is 0.232) tends to be larger than the mean value-weighted CAR (t-1 to t+1 mean is 0.032). Both have very large variations (the standard deviation is 0.309 for the equal-weighted and 0.094 for the value-weighted cases). That is, on average, the CARs around the merger announcement dates are not statistically different from zero.

The survival rate, summarized in Panel E, shows that more than 90% of acquiring firms survive one year after merger, and around 80% of them tend to be active two years after the completion of the merger. The survival rate is somewhat higher when both acquirors and targets are publicly traded (a mean value of 85%).

## Table 1. Summary Statistics

This table shows summary statistics for the variables used in this paper. Panel A shows the mean, standard deviation, and 5/50/95 percentiles of variables for Fama-French 12 industries (Fama and French (1997); https://mba.tuck.dartmouth.edu/pages/faculty/ken.french/Data_Library/det_12_ind_port.html) pairs by year using panel data from 1990 to 2021. *Unadjusted Distance* is the ratio of the out-of-sample mean squared errors of the acquiror's industry's production function using the target's industry's data to using the acquiror's industry's data. *TF Distance* is the ratio of the out-of-sample mean squared errors of the acquiror's industry's production function using the target's industry data to using the acquiror's industry data, but allowing adjustment of the production function at the last production stage (i.e., mapping from final production factors to production outcome) when using the target's industry's data. For both measures, we use a deep neural network for production functions that maps production input (deviation from the industry average of the natural logarithm of total assets, capital expenditures divided by total assets, short term debt divided by total assets, long term debt divided by total assets, employment divided by total assets, property plant and equipment divided by total assets, advertisement expenses divided by total assets, and R&D expenditure divided by total assets) to production output (deviation from the industry average of the natural logarithm of Tobin's Q for the baseline distance measure, and ROA (operating income divided by total assets) for the alternative distance measure). Both Unadjusted Distance and TF Distance are estimated in a cross-section of Compustat firms for each year. *Log(Unadjusted Distance)* is the natural logarithm of Unadjusted Distance. *Log(TF Distance)* is the natural logarithm of TF distance. *Number of M&A Deals* is the number of deals for each acquiror-target Fama French 12 industries pairs reported in SDC Platinum data. *Log(Number of M&A Deals)* is the natural logarithm of the number of M&A deals. Panel B shows variables used in the SDC Platinum database (SDC) deal level data from 1990 to 2021. There are three subpanels: The first subpanel shows all M&A deals; the second panel shows deals involving only publicly traded acquirors; and the last subpanel shows deals involving only publicly traded acquiror and publicly traded target pairs. We consider a (acquiror or target) firm to be publicly traded if the SDC CUSIP matches to the Compustat database. *Deal Completed* is one if the SDC records the M&A deal to be completed and zero otherwise. *Diversify* is an indicator for a diversifying merger (acquiror two digit SIC code differs from that of the target firm). *Hostile* is an indicator for a hostile merger as recorded in the SDC. *High Tech* is an indicator for both the acquiror and the target being in a high technology sector. *Tender Offer* is an indicator for when a tender offer is used as recorded in the SDC. *Stock Deal* is an indicator for when stock is (partly) used to finance the deal as reported in the SDC. *Relative Deal Size* is the deal value reported by the SDC divided by the market value of the acquiror's equity. *Firm size* is the natural logarithm of total assets from Compustat. *Tobin's Q* is total assets plus the market value of equity (total shares outstanding times year end stock price) minus the total value of common equity minus deferred taxes all divided by total assets from Compustat. *Book Leverage* is debt in current liabilities plus total long-term debt divided by total assets from Compustat. *Cash Flow to Assets* is operating income before depreciation minus total interest related expenses minus total income taxes minus capital expenditures divided by total assets from Compustat. Panel C shows annual (1990 to 2019) Fama-French 12 industries pair averages of 10-K product description based firm rival scores reported in the TNIC-3 industry database (Hoberg and Phillips (2010, 2016)). Panel D shows the daily cumulative abnormal returns around merger announcement day from t-1 to t, and from t-1 to t+1 (equal-weighted and value-weighted). Panel E shows post-merger survival which is 1 if the total asset exists in Compustat 1 or 2 years after the merger.

*Panel A. Industry Pair-Year Data (1990-2021)*

(1) Baseline Distance Measure: Fama French 12 Industry & log(Q) Output

| Variables | N | Mean | Std. Dev. | p5 | Median | p95 |
|---|---|---|---|---|---|---|
| Unadjsuted Distance | 4608 | 1.320 | 0.278 | 1.000 | 1.266 | 1.785 |
| TF Distance | 4608 | 1.285 | 0.214 | 1.074 | 1.219 | 1.715 |
| log(Unadjusted Distance) | 4608 | 0.261 | 0.173 | 0.000 | 0.236 | 0.580 |
| log(TF Distance) | 4608 | 0.239 | 0.149 | 0.071 | 0.198 | 0.540 |
| Number of M&A Deals | 4608 | 65 | 208 | 0 | 9 | 295 |
| log(Number of M&A Deals) | 4608 | 2.433 | 1.731 | 0.000 | 2.303 | 5.690 |

(2) Alternative Distance Measure: Fama-French 12 Industry & ROA Output

| Variables | N | Mean | Std. Dev. | p5 | Median | p95 |
|---|---|---|---|---|---|---|
| Unadjsuted Distance | 4608 | 1.552 | 0.581 | 1.000 | 1.402 | 2.539 |
| TF Distance | 4608 | 1.338 | 0.307 | 1.043 | 1.277 | 1.793 |
| log(Unadjusted Distance) | 4608 | 0.392 | 0.289 | 0.000 | 0.338 | 0.932 |
| log(TF Distance) | 4608 | 0.272 | 0.181 | 0.043 | 0.245 | 0.584 |

*(continued)*

**Table 1. Summary Statistics (continued)**

*Panel B. SDC Platinum Deal-Level Data (1990-2021)*
*Variables*
(1) All Deals

| Variables | N | Mean | Std. Dev. | p5 | Median | p95 |
|---|---|---|---|---|---|---|
| Deal Completed | 35613 | 0.892 | 0.310 | 0.000 | 1.000 | 1.000 |
| Diversify | 35613 | 0.481 | 0.500 | 0.000 | 0.000 | 1.000 |
| Hostile | 35613 | 0.009 | 0.096 | 0.000 | 0.000 | 0.000 |
| High Tech | 35613 | 0.214 | 0.410 | 0.000 | 0.000 | 1.000 |
| Tender Offer | 35613 | 0.050 | 0.217 | 0.000 | 0.000 | 0.000 |
| Stock Deal | 35613 | 0.026 | 0.159 | 0.000 | 0.000 | 0.000 |
| Relative Deal Size | 35613 | 4.074 | 257.060 | 0.097 | 1.000 | 1.389 |

(2) Public Acquirors

| Variables | N | Mean | Std. Dev. | p5 | Median | p95 |
|---|---|---|---|---|---|---|
| Deal Completed | 12365 | 0.917 | 0.275 | 0.000 | 1.000 | 1.000 |
| Diversify | 12365 | 0.384 | 0.486 | 0.000 | 0.000 | 1.000 |
| Hostile | 12365 | 0.009 | 0.093 | 0.000 | 0.000 | 0.000 |
| High Tech | 12365 | 0.340 | 0.474 | 0.000 | 0.000 | 1.000 |
| Tender Offer | 12365 | 0.054 | 0.226 | 0.000 | 0.000 | 1.000 |
| Stock Deal | 12365 | 0.027 | 0.163 | 0.000 | 0.000 | 0.000 |
| Relative Deal Size | 12365 | 1.500 | 27.288 | 0.190 | 1.000 | 1.439 |
| Firm Size (Acquiror) | 12365 | 6.562 | 2.541 | 2.332 | 6.642 | 10.505 |
| Tobin's Q (Acquiror) | 12365 | 5.751 | 152.891 | 0.942 | 1.545 | 7.827 |
| Book Leverage (Acquiror) | 12365 | 0.283 | 2.827 | 0.000 | 0.159 | 0.622 |
| Cash Flow to Asset (Acquiror) | 12365 | -18.076 | 774.862 | -0.013 | 0.000 | 0.001 |

(3) Both Public Acquiror & Public Target

| Variables | N | Mean | Std. Dev. | p5 | Median | p95 |
|---|---|---|---|---|---|---|
| Deal Completed | 3304 | 0.887 | 0.317 | 0.000 | 1.000 | 1.000 |
| Diversify | 3304 | 0.346 | 0.476 | 0.000 | 0.000 | 1.000 |
| Hostile | 3304 | 0.026 | 0.158 | 0.000 | 0.000 | 0.000 |
| High Tech | 3304 | 0.426 | 0.495 | 0.000 | 0.000 | 1.000 |
| Tender Offer | 3304 | 0.174 | 0.379 | 0.000 | 0.000 | 1.000 |
| Stock Deal | 3304 | 0.019 | 0.137 | 0.000 | 0.000 | 0.000 |
| Relative Deal Size | 3304 | 1.360 | 15.013 | 0.123 | 1.000 | 1.711 |
| Firm Size (Acquiror) | 3304 | 7.529 | 2.360 | 3.646 | 7.584 | 11.302 |
| Tobin's Q (Acquiror) | 3304 | 2.362 | 3.218 | 0.957 | 1.597 | 5.780 |
| Book Leverage (Acquiror) | 3304 | 0.231 | 0.206 | 0.000 | 0.198 | 0.601 |
| Cash Flow-to Asset (Acquiror) | 3304 | -0.015 | 0.179 | -0.001 | 0.000 | 0.001 |
| Firm Size (Target) | 3304 | 5.499 | 2.013 | 2.398 | 5.405 | 8.985 |
| Tobin's Q (Target) | 3304 | 4.774 | 156.313 | 0.806 | 1.351 | 5.464 |
| Book Leverage (Target) | 3304 | 0.235 | 0.348 | 0.000 | 0.160 | 0.676 |
| Cash Flow to Asset (Target) | 3304 | -0.014 | 0.185 | -0.018 | 0.000 | 0.002 |

*(continued)*

## 4. Results

**4.1 M&A Activities and Production Process Heterogeneity Across Industries**

Table 2 reports results from cross-sectional regressions of log(Number of M&A Deals) on our two functional distances, both of which are log-transformed and used in the RHS of the regressions. The table shows the negative point estimates for log(Unadjusted Distance) and log(TF Distance). The results hold throughout the years during our sample period from 1990 to 2021. Among the two measures, log(TF Distance) effects seem more robust both statistically and economically. These results are consistent with our **Propositions 2 and 3**.

In Table 3, we conduct pooled panel regressions using the same dependent variable, log(Number of M&A Deals), and our log-transformed functional distances on the RHS of the regressions. In the first two columns, we do not control for year and industry fixed effects, whereas we do control for them in the remaining columns of the table. In Column 1, with a one standard deviation increase in log(Unadjusted Distance), there is a 30% reduction in log(Number of M&A Deals) from its sample standard deviation. When log(TF Distance) is used as an alternative explanatory variable in Column 2, the economic significance is amplified to a 36% reduction in log(Number of M&A Deals). Both of these measures of production process heterogeneity between two of the Fama-French 12 industries seem to be able to significantly explain merger activities between them. They explain our dependent variable, log(Number of M&A Deals), at the 1% statistical significance level. Even after we incorporate year and industry fixed effects, both the statistical and economic significance of these variables are largely unchanged.

### Table 2. M&A Activities (Year-By-Year: 1990 - 2021)

This table shows the year-by-year relationship between M&A activities and industry distance measures from 1990 to 2021. The dependent variable is log(Number of M&A Deals). The key independent variables are log(Unadjusted Distance) in Columns (I) (estimate) and (II) (t-statistics), and log(TF Distance) in Columns (III) (estimate) and (IV) (t-statistics). Each regression has 144 observations (Fama-French 12 industries pairs) and includes intercepts (not shown in the table). The t-statistics are shown in square brackets. ***, **, and * denote significance at the 1%, 5%, and 10% level, respectively.

**Table 2 (continued)**

| Year | Log(Unadjusted Distance) | | log(TF Distance) | |
|---|---|---|---|---|
| | Estimate | t-stat | Estimate | t-stat |
| | (I) | (II) | (III) | (IV) |
| 1990 | -4.819*** | [-6.82] | -3.844*** | [-5.07] |
| 1991 | -3.007*** | [-4.59] | -4.264*** | [-4.57] |
| 1992 | -2.917*** | [-3.80] | -4.531*** | [-4.63] |
| 1993 | -2.258*** | [-2.99] | -6.706*** | [-6.27] |
| 1994 | -2.595*** | [-3.16] | -6.334*** | [-5.66] |
| 1995 | -1.438** | [-2.14] | -5.948*** | [-4.99] |
| 1996 | -4.134*** | [-4.73] | -6.277*** | [-6.07] |
| 1997 | -4.864*** | [-5.00] | -7.631*** | [-5.23] |
| 1998 | -5.420*** | [-5.99] | -7.839*** | [-6.53] |
| 1999 | -2.195*** | [-3.26] | -7.360*** | [-6.62] |
| 2000 | -5.134*** | [-5.53] | -5.492*** | [-5.01] |
| 2001 | -2.930*** | [-5.08] | -4.064*** | [-5.50] |
| 2002 | -7.227*** | [-7.37] | -4.425*** | [-4.70] |
| 2003 | -5.641*** | [-5.40] | -3.879*** | [-3.44] |
| 2004 | -5.051*** | [-5.27] | -4.682*** | [-4.75] |
| 2005 | -3.378*** | [-3.20] | -4.535*** | [-4.09] |
| 2006 | -3.545*** | [-3.71] | -4.015*** | [-4.21] |
| 2007 | -3.374*** | [-3.86] | -4.235*** | [-4.10] |
| 2008 | -2.278*** | [-4.50] | -2.172*** | [-3.73] |
| 2009 | -4.346*** | [-5.05] | -3.722*** | [-4.72] |
| 2010 | -3.577*** | [-4.11] | -3.741*** | [-4.41] |
| 2011 | -3.957*** | [-5.32] | -2.990*** | [-4.26] |
| 2012 | -3.770*** | [-4.41] | -3.693*** | [-4.34] |
| 2013 | -2.402*** | [-3.28] | -5.264*** | [-5.91] |
| 2014 | -2.937*** | [-3.81] | -4.018*** | [-4.95] |
| 2015 | -1.862** | [-2.27] | -3.244*** | [-3.37] |
| 2016 | -2.716*** | [-3.51] | -2.722*** | [-3.79] |
| 2017 | -0.984 | [-1.02] | -3.095*** | [-2.84] |
| 2018 | -2.709*** | [-2.76] | -4.032*** | [-4.02] |
| 2019 | -1.848** | [-2.27] | -4.189*** | [-4.54] |
| 2020 | -1.513* | [-1.91] | -3.537*** | [-4.02] |
| 2021 | -1.122 | [-1.24] | -5.251*** | [-4.53] |

**Table 3. M&A Activities (Industry Pair-Year Panel: 1990-2021)**

This table shows the relationship between M&A activities and industry distance measures in the Fama-French 12 industries pairs by year panel from 1990 to 2021. The dependent variable is log(Number of M&A Deals). The key independent variables are log(Unadjusted Distance) in Columns (I), (III), and (V), and are log(TF Distance) in Columns (II), (IV) and (VI). Each regression has 4608 observations (144 industry pairs times 32 years). Columns (I) and (II) have no fixed effects. Columns (III) and (IV) include year fixed effects. Columns (V) and (VI) include both year and industry fixed effects. The t-statistics are shown in square brackets and standard errors are clustered at the year level. ***, **, and * denote significance at the 1%, 5%, and 10% level, respectively.

**Table 3 (continued)**

| Dependent Variable | log(Number of M&A Deals) | | | | | |
|---|---|---|---|---|---|---|
| | (I) | (II) | (III) | (IV) | (V) | (VI) |
| log(Unadjusted Distance) | -2.962*** | | -3.000*** | | -4.620*** | |
| | [-13.85] | | [-13.76] | | [-18.97] | |
| log(TF Distance) | | -4.137*** | | -4.157*** | | -4.132*** |
| | | [-15.83] | | [-16.38] | | [-17.35] |
| Intercept | 3.206*** | 3.420*** | 3.216*** | 3.425*** | 3.279*** | 3.120*** |
| | [54.76] | [51.70] | [56.48] | [56.55] | [47.00] | [42.27] |
| Year FE | No | No | Yes | Yes | Yes | Yes |
| Industry FE | No | No | No | No | Yes | Yes |
| Observations | 4,608 | 4,608 | 4,608 | 4,608 | 4,608 | 4,608 |
| R-squared | 0.087 | 0.126 | 0.097 | 0.132 | 0.511 | 0.444 |

Now we turn our attention to the deal-level analysis. We run a probit regression with the Deal Completion indicator as our main dependent variable. We use log(Unadjusted Distance) and log(TF Distance) as our main explanatory variables. Control variables include the following deal characteristics (Diversify, Hostile, High Tech, Tender Offer, Stock Deal, Relative Deal Size) as the common set of controls for all three subsamples of the analysis (i.e., All Deals, Public Acquiror Deals, Public Acquiror and Target Deals). When we observe the firm characteristics (Public Acquiror Deals, Public Acquiror and Target Deals), we further control for Tobin's Q, Book Leverage, and Cash Flow to Assets of acquiror and target whenever they are available. The results are reported in Table 4; Panel A reports the marginal effect on the probit propensity of each explanatory variable, and Panel B reports their point estimates.

### Table 4. Deals Completed (SDC Platinum Deal Level: 1990-2021)

This table shows the relationship between the likelihood of M&A deal completion and industry distance measures from 1990 to 2021. The dependent variable is an indicator for M&A deal completion, which is one if the deal is completed and zero otherwise. The key independent variables are log(Unadjusted Distance) in Columns (I), (III), (V), and (VII), and are log(TF Distance) in Columns (II), (IV), (VI), and (VIII). Columns (I) to (IV) are probit regressions for deals involving both public and private firms (acquiror and target). Columns (V) and (VI) are probit regressions for deals involving publicly traded acquirors. Columns (VII) and (VIII) are probit regressions for deals involving publicly traded acquirors and publicly traded targets. All regressions are controlled for Diversify, Hostile, High Tech, Tender Offer, Stock Deal, and Relative Deal Size. Additionally, Columns (V) and (VI) control for acquirors' Firm Size, Tobin's Q, Book Leverage, and Cash Flow to Assets ratio, and year and industry fixed effects; Columns (VII) and (VIII) control for both acquirors' and targets' Firm Size, Tobin's Q, Book Leverage, and Cash Flow to Assets ratio, as well as year and industry fixed effects. The t-statistics are shown in square brackets and standard errors are clustered at the year level. ***, **, and * denote significance at the 1%, 5%, and 10% level, respectively.

# Table 4 (continued)

Panel A. Marginal Effects

| Dependent Variable | Indicator for Deal Completion | | | | | | | |
|---|---|---|---|---|---|---|---|---|
| | All Deals | | | | Public Acquiror | | Acquiror & Target Public | |
| | (I) | (II) | (III) | (IV) | (V) | (VI) | (VII) | (VIII) |
| log(Unadjusted Distance) | -0.071*** | | -0.056*** | | -0.041* | | -0.073 | |
| | [-4.26] | | [-3.47] | | [-1.89] | | [-1.19] | |
| log(TF Distance) | | -0.077*** | | -0.074*** | | -0.066** | | -0.141** |
| | | [-3.38] | | [-3.73] | | [-2.45] | | [-1.98] |
| Diversify | -0.022*** | -0.035*** | -0.021*** | -0.031*** | -0.007 | -0.012** | -0.019 | -0.026* |
| | [-4.26] | [-6.59] | [-3.90] | [-5.63] | [-1.02] | [-1.98] | [-1.11] | [-1.95] |
| Hostile | -0.393*** | -0.391*** | -0.402*** | -0.400*** | -0.319*** | -0.319*** | -0.331*** | -0.331*** |
| | [-13.55] | [-13.53] | [-15.75] | [-15.63] | [-18.67] | [-18.70] | [-13.78] | [-13.81] |
| High Tech | 0.018*** | 0.018*** | 0.023*** | 0.022*** | -0.002 | -0.002 | -0.014 | -0.015 |
| | [2.98] | [2.87] | [3.81] | [3.66] | [-0.27] | [-0.30] | [-0.77] | [-0.82] |
| Tender Offer | 0.041*** | 0.041*** | 0.038*** | 0.038*** | 0.028 | 0.028 | 0.076*** | 0.076*** |
| | [3.70] | [3.69] | [3.74] | [3.74] | [1.61] | [1.59] | [3.54] | [3.53] |
| Stock Deal | 0.032 | 0.029 | 0.055*** | 0.053*** | 0.033 | 0.032 | 0.030 | 0.031 |
| | [1.62] | [1.39] | [3.67] | [3.55] | [1.30] | [1.28] | [0.48] | [0.48] |
| Relative Deal Size | -0.000 | -0.000 | -0.000 | -0.000 | -0.000 | -0.000 | -0.008** | -0.009** |
| | [-0.53] | [-0.54] | [-0.47] | [-0.48] | [-1.56] | [-1.57] | [-2.09] | [-2.07] |
| Firm Size (Acquiror) | | | | | 0.007*** | 0.007*** | 0.022*** | 0.022*** |
| | | | | | [4.02] | [4.22] | [4.81] | [4.84] |
| Tobin's Q (Acquiror) | | | | | 0.000 | 0.000 | 0.001 | 0.001 |
| | | | | | [1.43] | [1.45] | [0.75] | [0.62] |
| Book Leverage (Acquiror) | | | | | -0.001 | -0.001 | -0.080*** | -0.080*** |
| | | | | | [-1.07] | [-1.04] | [-3.63] | [-3.61] |
| Cash Flow-to Asset (Acquiror) | | | | | 0.000 | 0.000 | -0.083 | -0.084 |
| | | | | | [0.60] | [0.62] | [-1.41] | [-1.43] |
| Firm Size (Target) | | | | | | | -0.017*** | -0.017*** |
| | | | | | | | [-3.09] | [-3.01] |
| Tobin's Q (Target) | | | | | | | 0.000 | 0.000 |
| | | | | | | | [0.09] | [0.28] |
| Book Leverage (Target) | | | | | | | -0.003 | -0.002 |
| | | | | | | | [-0.17] | [-0.08] |
| Cash Flow to Asset (Target) | | | | | | | -0.005 | -0.002 |
| | | | | | | | [-0.15] | [-0.07] |
| Model | Probit | Probit | Probit | Probit | Probit | Probit | Probit | Probit |
| Public Acquiror | | | | | Yes | Yes | Yes | Yes |
| Public Target | | | | | | | Yes | Yes |
| Year FE | No | No | Yes | Yes | Yes | Yes | Yes | Yes |
| Industry FE | No | No | No | No | Yes | Yes | Yes | Yes |
| Pseudo R2 | 0.046 | 0.045 | 0.060 | 0.060 | 0.081 | 0.080 | 0.157 | 0.157 |
| Observations | 35,613 | 35,613 | 35,613 | 35,613 | 12,365 | 12,365 | 3,304 | 3,304 |

We focus on the marginal effects of our main explanatory variables, reported in Panel A. In Column 1 of Panel A, we focus on All Deals and find there is a 7.1% reduction in the M&A completion likelihood for a marginal increase in log(Unadjusted Distance). A similar magnitude of negative effect from log(TF Distance) is found in Column 2, which corresponds to a -7.7% reduction in M&A completion likelihood. Both effects are statistically significant at the 1% level. The results are largely robust to more control variables as we limit the sample to public acquirors and targets (Columns 5, 6, 7, and

8) and include year and industry fixed effects (Columns 3, 4, 5, 6, 7 and 8). Below, we further report the coefficients of all the RHS variables in our probit regressions.

## Table 4 (continued)

Panel B. Coefficients

| Dependent Variable | Indicator for Deal Completion | | | | | | | |
|---|---|---|---|---|---|---|---|---|
| | All Deals | | | | Public Acquiror | | Acquiror & Target Public | |
| | (I) | (II) | (III) | (IV) | (V) | (VI) | (VII) | (VIII) |
| log(Unadjusted Distance) | -0.401*** | | -0.322*** | | -0.316** | | -0.563 | |
| | [-4.34] | | [-3.45] | | [-2.02] | | [-1.40] | |
| log(TF Distance) | | -0.438*** | | -0.433*** | | -0.460** | | -1.013** |
| | | [-3.38] | | [-3.75] | | [-2.30] | | [-2.15] |
| Diversify | -0.127*** | -0.198*** | -0.122*** | -0.176*** | -0.045 | -0.084** | -0.099 | -0.154** |
| | [-4.18] | [-6.65] | [-3.87] | [-5.54] | [-0.97] | [-2.07] | [-0.98] | [-2.01] |
| Hostile | -2.231*** | -2.219*** | -2.312*** | -2.303*** | -2.281*** | -2.282*** | -2.003*** | -2.004*** |
| | [-14.03] | [-14.00] | [-14.70] | [-14.62] | [-17.87] | [-17.89] | [-13.19] | [-13.25] |
| High Tech | 0.104*** | 0.103*** | 0.131*** | 0.130*** | -0.008 | -0.009 | -0.080 | -0.088 |
| | [3.07] | [2.95] | [3.83] | [3.67] | [-0.14] | [-0.15] | [-0.70] | [-0.75] |
| Tender Offer | 0.231*** | 0.231*** | 0.217*** | 0.216*** | 0.220* | 0.216* | 0.490*** | 0.484*** |
| | [3.92] | [3.92] | [3.75] | [3.74] | [1.76] | [1.74] | [3.73] | [3.72] |
| Stock Deal | 0.181 | 0.166 | 0.316*** | 0.306*** | 0.235 | 0.225 | 0.172 | 0.175 |
| | [1.58] | [1.35] | [3.67] | [3.55] | [1.32] | [1.29] | [0.44] | [0.45] |
| Relative Deal Size | -0.000 | -0.000 | -0.000 | -0.000 | -0.001 | -0.001 | -0.048* | -0.048* |
| | [-0.53] | [-0.54] | [-0.47] | [-0.48] | [-1.60] | [-1.61] | [-1.79] | [-1.78] |
| Firm Size (Acquiror) | | | | | 0.047*** | 0.049*** | 0.122*** | 0.121*** |
| | | | | | [4.05] | [4.31] | [3.85] | [3.88] |
| Tobin's Q (Acquiror) | | | | | 0.002 | 0.002 | 0.004 | 0.003 |
| | | | | | [1.45] | [1.46] | [0.56] | [0.40] |
| Book Leverage (Acquiror) | | | | | -0.005 | -0.005 | -0.451*** | -0.460*** |
| | | | | | [-0.94] | [-0.91] | [-3.24] | [-3.25] |
| Cash Flow-to Asset (Acquiror) | | | | | 0.000 | 0.000 | -0.441 | -0.446 |
| | | | | | [0.70] | [0.71] | [-1.29] | [-1.32] |
| Firm Size (Target) | | | | | | | -0.089** | -0.086** |
| | | | | | | | [-2.36] | [-2.24] |
| Tobin's Q (Target) | | | | | | | 0.000 | 0.000 |
| | | | | | | | [0.13] | [0.35] |
| Book Leverage (Target) | | | | | | | -0.042 | -0.032 |
| | | | | | | | [-0.37] | [-0.28] |
| Cash Flow to Asset (Target) | | | | | | | -0.087 | -0.069 |
| | | | | | | | [-0.41] | [-0.33] |
| Intercept | 1.358*** | 1.396*** | 1.256*** | 1.334*** | 1.246*** | 1.346*** | 1.142*** | 1.362*** |
| | [34.40] | [31.47] | [61.62] | [40.38] | [12.15] | [12.91] | [4.85] | [5.30] |
| Model | Probit | Probit | Probit | Probit | Probit | Probit | Probit | Probit |
| Public Acquiror | | | | | Yes | Yes | Yes | Yes |
| Public Target | | | | | | | Yes | Yes |
| Year FE | No | No | Yes | Yes | Yes | Yes | Yes | Yes |
| Industry FE | No | No | No | No | Yes | Yes | Yes | Yes |
| Pseudo R2 | 0.046 | 0.045 | 0.060 | 0.060 | 0.081 | 0.080 | 0.157 | 0.157 |
| Observations | 35,613 | 35,613 | 35,613 | 35,613 | 12,365 | 12,365 | 3,304 | 3,304 |

Overall, the results reported in Tables 2, 3, and 4 support our **Propositions 2** and **3**, that M&As are less likely when the acquiror and target are from industries with large differences in their production decision processes. Among the two functional distances, TF Distance seems to capture more salient differences in the production functions between the acquiror's and the target's industries.

## 4.2 Is Production Process Heterogeneity Related to Text-Based Network Industry Classifications (TNIC) by Hoberg and Phillips (2010, 2016)?

Hoberg and Phillips (2010, 2016) pioneered the way we classify industries. They focus on each firm's text-based product description in their annual filings, i.e., their 10-Ks. Based on this description of the end product markets in which each firm operates, Hoberg and Phillips re-group firms into more relevant product peers/rivals. These are TNIC scores. The higher the TNIC scores, the more similar two firms' product characteristics are. Hoberg and Phillips provide SIC3 digit-based TNIC scores (TNIC-3) in their data library, which we download and use in our analysis here.

By the definition of TNIC3 score, the higher the score is, the smaller the functional distance between acquiror and target. Below we test the contemporaneous correlations between TNIC3 score and log(Unadjusted Distance) and log(TF Distance). As seen in Table 5, Panel A, TNIC3 scores are negatively correlated with our functional distance measures. The magnitude of the correlation ranges from 7% to 10%, all of which are statistically significant at the 10% level.

In Panel B of Table 5, we further conduct predictive regressions between these variables. In the first two columns, we confirm the contemporaneous correlations between TNIC3 scores and our functional distances in panel regressions with year and industry fixed effects further controlled. They are statistically significant at the 1% level.

## Table 5. Hoberg-Phillips TNIC3 Score (1990-2019)

This table shows the relationship between Hoberg-Phillips (2010, 2016) TNIC-3 database firm rival scores and industry distance measures from 1990 to 2019. A TNIC3 score is an annual (1990 to 2019) Fama-French 12 industries pair average of firm rival scores based on 10-K product descriptions, reported in the TNIC-3 industry database (Hoberg and Phillips (2010, 2016)). Panel A shows correlations among TNIC3 Score, log(Unadjusted Distance), and log(TF Distance). * indicates a significance level of correlation coefficients at the 1% level or better. Panel B shows panel regressions of TNIC3 Score on log(Unadjusted Distance) in Column (I), log(TF Distance) in Column (II), lag of log(Unadjusted Distance) in Column (IIII), lag of log(TF Distance) in Column (IV), log(Unadjusted Distance) on lag of TNIC3 Score in Column (V), and log(TF Distance) on lag of TNIC3 Score in Column (VI). All regressions control for intercept, year and industry fixed effects. Panel C shows the relationship between M&A activity (log of Number of M&A Deals) and both TNIC3 scores and industry distance measures. The t-statistics are shown in square brackets and standard errors are clustered at the year level. ***, **, and * denote significance at the 1%, 5%, and 10% level, respectively.

**Panel A. Correlations**

|  | TNIC3 Score | log(Unadjusted Distance) |
|---|---|---|
| log(Unadjusted Distance) | -0.1046* |  |
| log(TF Distance) | -0.0744* | 0.5370* |

# Table 5 (continued)

**Panel B. TNIC3 Score Regression**

| Dependent Variable | TNIC3 Score | | TNIC3 Score | | log(Unadjusted Distance) | log(TF Distance) |
|---|---|---|---|---|---|---|
| | (I) | (II) | (III) | (IV) | (V) | (VI) |
| log(Unadjusted Distance(t)) | -0.017*** | | | | | |
| | [-9.04] | | | | | |
| log(Unadjusted Distance(t-1)) | | | -0.016*** | | | |
| | | | [-7.81] | | | |
| log(TF Distance(t)) | | -0.012*** | | | | |
| | | [-5.90] | | | | |
| log(TF Distance(t-1)) | | | | -0.011*** | | |
| | | | | [-5.09] | | |
| TNIC3 Score(t-1) | | | | | -1.119*** | -0.689*** |
| | | | | | [-3.30] | [-3.79] |
| Intercept | 0.032*** | 0.031*** | 0.032*** | 0.031*** | 0.278*** | 0.255*** |
| | [39.83] | [27.69] | [33.37] | [25.69] | [20.04] | [39.36] |
| | | | | | | |
| Year FE | Yes | Yes | Yes | Yes | Yes | Yes |
| Industry FE | Yes | Yes | Yes | Yes | Yes | Yes |
| Observations | 4,300 | 4,300 | 4,138 | 4,138 | 4,138 | 4,138 |
| R-squared | 0.075 | 0.066 | 0.074 | 0.065 | 0.129 | 0.037 |

In Columns 3 and 4 of Panel B, we further check the one year lead-lag relationship between TNIC3 score (the LHS variable) and our functional distances on the RHS, log(Unadjusted Distance) in Column 3 and log(TF Distance) in Column 4. There we find significantly negative coefficients for both of our distance measures, predicting the TNIC3 score in the following year. However, when we repeat similar analyses using log(Unadjusted Distance) as the LHS variable while we have the lagged TNIC3 score on the RHS, we find similar predictive powers for the TNIC3 score on log(Unadjusted Distance). We also find similar predictive powers for the TNIC3 score on log(TF Distance). All the results are statistically significant at the 1% level.

Lastly, in Panel C of Table 5, we run horse race regressions between TNIC3 scores and our functional distances with log(Number of M&A Deals) as the main explanatory variable. In the panel regressions with and without year and industry fixed effects, we find that both measures are complementary and jointly capture the dimensions that are relevant to merger synergy. All of the variables significantly explain log(Number of M&A Deals). Their effects are statistically significant at the 1% level. The effects are also economically meaningful. For example, in Column 6 of Panel C, we find that a one standard deviation increase in the TNIC3 score explains a 13% increase in merger intensity, while a one standard deviation increase in log(TF Distance) explains a 35% reduction in merger intensity.

## Table 5 (continued)

**Panel C. log(Number of M&A Deals) Regressions**

| Dependent Variable | log(Number of M&A Deals) | | | | | |
|---|---|---|---|---|---|---|
| | (I) | (II) | (III) | (IV) | (V) | (VI) |
| TNIC3 Score | 12.132*** | 12.467** | 8.596** | 12.547*** | 12.811*** | 10.910*** |
| | [2.81] | [2.73] | [2.67] | [2.89] | [2.80] | [2.97] |
| log(Unadjusted Distance) | -2.938*** | -2.963*** | -4.564*** | | | |
| | [-13.79] | [-13.79] | [-17.09] | | | |
| log(TF Distance) | | | | -4.000*** | -3.997*** | -4.019*** |
| | | | | [-14.37] | [-14.87] | [-16.15] |
| Intercept | 2.835*** | 2.831*** | 3.027*** | 3.008*** | 2.999*** | 2.789*** |
| | [18.09] | [17.20] | [21.73] | [18.84] | [18.01] | [19.13] |
| | | | | | | |
| Year FE | No | Yes | Yes | No | Yes | Yes |
| Industry FE | No | No | Yes | No | No | Yes |
| Observations | 4,300 | 4,300 | 4,300 | 4,300 | 4,300 | 4,300 |
| R-squared | 0.117 | 0.128 | 0.521 | 0.150 | 0.156 | 0.454 |

### 4.3 Do Unadjusted Distance and TF Distance Identify the Underlying Production Factors and Weights Differences?

As we discuss in the corollaries in Section 2, careful inspection of the joint condition of Unadjusted Distance and TF Distance appears to explain whether two industries are drastically different in both production factors and weights or share common production factors yet different weights associated with those factors. To test **Corollaries 1**, **2**, and **3** jointly, we regress log(Number of M&A Deals) on log(Unadjusted Distance), log(TF Distance), and the interaction between them. We carefully disentangle the highly correlated nature of the two distances and avoid any multicollinearity issues. We orthogonalize log(TF Distance) from log(Unadjusted Distance) and use the residual from the orthogonalization regression as our main independent variable, which is denoted by log(TF Distance) Residual.

Results are reported in Table 6. We first find that log(Unadjusted Distance) significantly explains the reduction in M&A intensity between two industries with a high Unadjusted Distance. Then we find a significantly negative point estimate of the interaction term between log(Unadjusted Distance) and log(TF Distance) Residual. This result indicates that two industries with a high log(Unadjusted Distance) and a low log(TF Distance) are less likely to suffer from a high cost of business integration than industries with both a high log(Unadjusted Distance) and a high log(TF Distance).

These findings are consistent with our corollaries, reconfirming our theoretical intuitions obtained from a linear stylized model of our deep neural network.

## Table 6. M&A Activities (Industry Pair-Year Panel: 1990-2021): Interaction Between Distance Measures

This table shows the relationship between M&A activities and the interaction of industry distance measures using the Fama-French 12 industries pairs by year panel from 1990 to 2021. The dependent variable is log(Number of M&A Deals). The key independent variables are log(Unadjusted Distance), residual of log(TF Distance), and their interactions. Due to the high correlation between log(Unadjusted Distance) and log(TF Distance), we orthogonalize log(TF Distance) by regressing log(TF Distance) on log(Unadjusted Distance) and year and industry fixed effects, and take residuals. Each regression has 4608 observations (144 industry pairs times 32 years). Column (I) has no fixed effects. Column (II) includes year fixed effects. Column (III) includes both year and industry fixed effects. The t-statistics are shown in square brackets and standard errors are clustered at the year level. ***, **, and * denote significance at the 1%, 5%, and 10% level, respectively.

| Dependent Variable | log(Number of M&A Deals) | | |
| --- | --- | --- | --- |
|  | (I) | (II) | (III) |
| log(Unadjusted Distance) | -2.832*** | -2.872*** | -4.519*** |
|  | [-14.11] | [-14.07] | [-18.01] |
| log(TF Distance) Residual | 0.365 | 0.312 | -0.571 |
|  | [0.65] | [0.57] | [-1.20] |
| log(Unadjusted Distance) | -7.097*** | -6.929*** | -4.140** |
| x log(TF Distance) Residual | [-3.03] | [-3.00] | [-2.10] |
| Intercept | 3.172*** | 3.183*** | 3.260*** |
|  | [53.79] | [59.73] | [43.60] |
| Year FE | No | Yes | Yes |
| Industry FE | No | No | Yes |
| Observations | 4,608 | 4,608 | 4,608 |
| R-squared | 0.118 | 0.126 | 0.533 |

**4.4 Robustness to the Alternative Production Outcome Specification (ROA-Based Distances)**

Now we carry out several further tests to document the robustness of our results. The analysis mainly focuses on the potential criticism that Tobin's Q as the outcome of the production function might not be valid, and other alternative outcome variables are more suitable for our product process estimation. Hence, we consider ROA as an alternative production outcome in our deep learning. Using the new set of distances obtained from this alternative specification, we show the robustness of our main results, reported in Tables 3, 4, and 5. The robustness test results are reported in Table 7, Panel A, Table 7, Panel B, and Table 8. In all these replicating regressions, we find that our results are largely unaffected by this specification change in our deep neural network models.

# Table 7. An Alternative Specification for Industry Distance: M&A Activities (Industry Pair-Year Panel: 1990-2021)

This table repeats the empirical tests from Table 3 (Panel A) and Table 4 (Panel B) using an alternative measure of industry distance; we use ROA instead of log(Q) when estimating industry production functions, which are used for estimating industry distances. The key independent variables are the ROA-based log(Unadjusted Distance) and ROA-based log(TF Distance). The t-statistics are shown in square brackets and standard errors are clustered at the year level. ***, **, and * denote significance at the 1%, 5%, and 10% level, respectively.

**Panel A. FF12 Industry Pair-Year Panel**

| Dependent Variable | log(Number of M&A Deals) | | | | | |
|---|---|---|---|---|---|---|
| | (I) | (II) | (III) | (IV) | (V) | (VI) |
| ROA-Based log(Unadjusted Distance) | -1.860*** | | -1.927*** | | -2.618*** | |
| | [-17.84] | | [-18.84] | | [-20.95] | |
| ROA-Based log(TF Distance) | | -2.894*** | | -3.036*** | | -2.992*** |
| | | [-8.42] | | [-8.92] | | [-9.70] |
| Intercept | 3.161*** | 3.221*** | 3.187*** | 3.260*** | 2.996*** | 2.942*** |
| | [60.82] | [36.48] | [79.56] | [35.17] | [42.56] | [30.03] |
| | | | | | | |
| Year FE | No | No | Yes | Yes | Yes | Yes |
| Industry FE | No | No | No | No | Yes | Yes |
| Observations | 4,608 | 4,608 | 4,608 | 4,608 | 4,608 | 4,608 |
| R-squared | 0.097 | 0.091 | 0.109 | 0.102 | 0.487 | 0.414 |

# Table 7 (continued)

**Panel B. SDC Platinum M&A Data (Dependent Variable: Deal Completed Indicator)**

| Dependent Variable | Indicator for Deal Completion | | | | | | | |
|---|---|---|---|---|---|---|---|---|
| | All Deals | | | | Public Acquiror | | Acquiror & Target Public | |
| | (I) | (II) | (III) | (IV) | (V) | (VI) | (VII) | (VIII) |
| ROA-Based log(Unadjusted Distance) | -0.257*** | | -0.219*** | | -0.268** | | -0.306 | |
| | [-4.35] | | [-3.32] | | [-2.48] | | [-1.39] | |
| ROA-Based log(TF-Distance) | | -0.323*** | | -0.278*** | | -0.297 | | -0.918*** |
| | | [-4.23] | | [-3.57] | | [-1.52] | | [-2.79] |
| Diversify | -0.142*** | -0.189*** | -0.130*** | -0.170*** | -0.044 | -0.088* | -0.105 | -0.108 |
| | [-4.88] | [-6.38] | [-4.44] | [-5.41] | [-0.87] | [-1.96] | [-1.09] | [-1.30] |
| Hostile | -2.233*** | -2.224*** | -2.311*** | -2.304*** | -2.296*** | -2.295*** | -2.013*** | -2.004*** |
| | [-14.08] | [-13.92] | [-14.76] | [-14.62] | [-18.18] | [-18.22] | [-12.90] | [-13.03] |
| High Tech | 0.109*** | 0.112*** | 0.135*** | 0.138*** | -0.025 | -0.015 | -0.082 | -0.089 |
| | [3.25] | [3.35] | [3.91] | [4.04] | [-0.44] | [-0.26] | [-0.75] | [-0.80] |
| Tender Offer | 0.233*** | 0.232*** | 0.218*** | 0.218*** | 0.227* | 0.225* | 0.501*** | 0.501*** |
| | [3.98] | [3.95] | [3.81] | [3.78] | [1.87] | [1.85] | [3.83] | [3.83] |
| Stock Deal | 0.162 | 0.161 | 0.304*** | 0.302*** | 0.221 | 0.218 | 0.181 | 0.173 |
| | [1.34] | [1.32] | [3.53] | [3.50] | [1.27] | [1.25] | [0.46] | [0.45] |
| Relative Deal Size | -0.000 | -0.000 | -0.000 | -0.000 | -0.001 | -0.001 | -0.051** | -0.052** |
| | [-0.53] | [-0.53] | [-0.47] | [-0.47] | [-1.61] | [-1.62] | [-2.13] | [-2.17] |
| Firm Size (Acquiror) | | | | | 0.048*** | 0.048*** | 0.126*** | 0.126*** |
| | | | | | [4.09] | [4.16] | [4.00] | [4.02] |
| Tobin's Q (Acquiror) | | | | | 0.002 | 0.002 | 0.004 | 0.004 |
| | | | | | [1.50] | [1.51] | [0.62] | [0.63] |
| Book Leverage (Acquiror) | | | | | -0.006 | -0.005 | -0.450*** | -0.451*** |
| | | | | | [-1.07] | [-1.03] | [-3.28] | [-3.28] |
| Cash Flow-to Asset (Acquiror) | | | | | 0.000 | 0.000 | -0.447 | -0.473 |
| | | | | | [0.67] | [0.69] | [-1.30] | [-1.24] |
| Firm Size (Target) | | | | | | | -0.087** | -0.087** |
| | | | | | | | [-2.32] | [-2.33] |
| Tobin's Q (Target) | | | | | | | 0.000 | 0.000 |
| | | | | | | | [0.21] | [0.22] |
| Book Leverage (Target) | | | | | | | -0.033 | -0.017 |
| | | | | | | | [-0.28] | [-0.14] |
| Cash Flow to Asset (Target) | | | | | | | -0.066 | -0.049 |
| | | | | | | | [-0.31] | [-0.24] |
| Intercept | 1.357*** | 1.379*** | 1.247*** | 1.287*** | 1.204*** | 1.267*** | 1.082*** | 1.268*** |
| | [34.21] | [33.71] | [65.38] | [52.12] | [11.82] | [10.46] | [4.76] | [5.44] |
| Model | Probit | Probit | Probit | Probit | Probit | Probit | Probit | Probit |
| Public Acquiror | | | | | Yes | Yes | Yes | Yes |
| Public Target | | | | | | | Yes | Yes |
| Year FE | No | No | Yes | Yes | Yes | Yes | Yes | Yes |
| Industry FE | No | No | No | No | Yes | Yes | Yes | Yes |
| Observations | 35,615 | 35,615 | 35,615 | 35,615 | 12,378 | 12,378 | 3,318 | 3,318 |

# Table 8. An Alternative Specification for Industry Distance: Hoberg-Phillips TNIC3 Score (1990-2019)

This table repeats the empirical tests from Table 5 using an alternative measure of industry distance; we use ROA instead of log(Q) when estimating industry production functions, which are used for estimating industry distances. The key independent variables are the ROA-based log(Unadjusted Distance) and the ROA-based log(TF Distance). Panel A shows correlations among TNIC3 Score, ROA-based log(Unadjusted Distance), and ROA-based log(TF Distance). * indicates a significance level of correlation coefficients at the 1% level or better. Panel B shows panel regressions of TNIC3 Score on ROA-based log(Unadjusted Distance) in Column (I) and ROA-based log(TF Distance) in Column (II). Both regressions control for intercept and year and industry fixed effects. Panel C shows the relationship between M&A activity (log of Number of M&A Deals) and both TNIC3 score and industry distance measures. The t-statistics are shown in square brackets and standard errors are clustered at the year level. ***, **, and * denote significance at the 1%, 5%, and 10% level, respectively.

**Panel A. Correlations**

|  | TNIC3 Score | ROA-Based log(Unadjusted Distance) |
|---|---|---|
| ROA-Based log(Unadjusted Distance) | -0.0717* |  |
| ROA-Based log(TF Distance) | -0.0489* | 0.5902* |

**Panel B. Panel Regression**

| Dependent Variable | TNIC3 Score | TNIC3 Score | TNIC3 Score | TNIC3 Score | log(Unadjusted Distance) | log(TF Distance) |
|---|---|---|---|---|---|---|
|  | (I) | (II) | (III) | (IV) | (V) | (VI) |
| ROA-Based log(Unadjusted Distance(t)) | -0.010*** |  |  |  |  |  |
|  | [-6.24] |  |  |  |  |  |
| ROA-Based log(Unadjusted Distance(t-1)) |  |  | -0.009*** |  |  |  |
|  |  |  | [-6.91] |  |  |  |
| ROA-Based log(TF Distance(t)) |  | -0.007*** |  |  |  |  |
|  |  | [-4.35] |  |  |  |  |
| ROA-Based log(TF Distance(t-1)) |  |  |  | -0.006** |  |  |
|  |  |  |  | [-2.61] |  |  |
| TNIC3 Score(t-1) |  |  |  |  | -1.695** | -0.404* |
|  |  |  |  |  | [-2.76] | [-1.78] |
| Intercept | 0.032*** | 0.030*** | 0.031*** | 0.030*** | 0.378*** | 0.282*** |
|  | [26.31] | [28.29] | [26.22] | [24.35] | [13.88] | [29.22] |
|  |  |  |  |  |  |  |
| Year FE | Yes | Yes | Yes | Yes | Yes | Yes |
| Industry FE | Yes | Yes | Yes | Yes | Yes | Yes |
| Observations | 4,300 | 4,300 | 4,138 | 4,138 | 4,138 | 4,138 |
| R-squared | 0.073 | 0.062 | 0.072 | 0.062 | 0.160 | 0.072 |

# Table 8 (continued)

**Panel C. log(Number of M&A Deals) Regression**

| Dependent Variable | log(Number of M&A Deals) | | | | | |
|---|---|---|---|---|---|---|
| | (I) | (II) | (III) | (IV) | (V) | (VI) |
| TNIC3 Score | 12.868*** | 13.104*** | 9.203*** | 13.502*** | 13.733*** | 11.929*** |
| | [3.02] | [2.92] | [2.94] | [2.93] | [2.85] | [3.02] |
| ROA-Based log(Unadjusted Distance) | -1.829*** | -1.894*** | -2.588*** | | | |
| | [-18.73] | [-19.92] | [-20.30] | | | |
| ROA-Based log(TF Distance) | | | | -2.758*** | -2.882*** | -2.867*** |
| | | | | [-7.81] | [-8.22] | [-8.96] |
| Intercept | 2.762*** | 2.780*** | 2.735*** | 2.777*** | 2.804*** | 2.579*** |
| | [19.42] | [19.10] | [20.57] | [16.87] | [15.82] | [16.00] |
| Year FE | | | | | | |
| Industry FE | | | | | | |
| Observations | 4,300 | 4,300 | 4,300 | 4,300 | 4,300 | 4,300 |
| R-squared | 0.124 | 0.137 | 0.495 | 0.115 | 0.127 | 0.423 |

## 4.5 Other M&A Performance Consequences

In this section, we further carry out additional tests that show the value of synergy, which is expected to be higher for mergers of firms in industries with minimal distances in their production decision processes. We posit that the expected synergy is more likely to be captured by stock return reactions around the deal announcement date. We also expect a greater chance of long-term survival of the combined organization if the acquiror smoothly integrates the target's main production technologies and generates the expected synergy ex-post. Below we discuss each of these two additional analyses and their results.

### 4.5.1 Announcement Effects

For each deal announcement, we construct (-1, 0)-day CAR or (-1, +1)-day CAR using a simple market model. Abnormal returns from the market model are cumulative over the two event test window. They serve as the main dependent variable in our announcement effect regressions.

The results are reported in Table 9. We run the linear panel regression using log(Unadjusted Distance) or log(TF Distance) as the main RHS variable. We control for the same list of deal characteristics and acquiror and target firm characteristics as in our previous deal-level analysis, Table 4.

In Column 1 of Table 9, the point estimate of -0.155 implies a -3% worse stock market reaction upon the announcement of a deal if the acquiror and target show a significant distance in their underlying production processes. In Column 3 of the same table, we confirm again a -3% worse stock market reaction during the 3-day window around the deal announcement date. These are results associated with log(Unadjusted distance) as our key independent variable. When we use log(TF Distance) instead, we find a 5% worse stock market reaction around the same 3-day window. Overall, investors carefully examine the potential integration cost of a merger and reflect their beliefs in stock prices during the short-term event window.

### Table 9. M&A Announcement Effect (SDC Platinum Deal Level: 1990-2021)

This table shows the relationship between M&A deal announcements and industry distance measures from 1990 to 2021 for deals involving both public and private firms (acquiror and target). The dependent variable is daily cumulative abnormal returns around announcement dates from t-1 to t (Columns (I) and (II)) or t-1 to t+1 (Columns (III) to (VI)). Columns (I) to (IV) equally weight acquiror and target cumulative abnormal returns. Columns (V) and (VI) value weight acquiror and target cumulative abnormal returns. The key independent variables are log(Unadjusted Distance) in Columns (I), (III), and (V) and log(TF Distance) in Columns (II), (IV), and (VI). All regressions are controlled for indicators for Merger Completed, Diversify, Hostile, High Tech, Tender Offer, Stock Deal, and Relative Deal Size, acquirors' & targets' firm size, Tobin's Q, Book Leverage, and Cash Flow to Asset ratio, and year and industry fixed effects. The t-statistics are shown in square brackets and standard errors are clustered at the year level. ***, **, and * denote significance at the 1%, 5%, and 10% level, respectively.

## Table 9 (continued)

| Dependent Variable | CAR(t-1 to t) | | CAR(t-1 to t+1) | | CAR(t-1 to t+1) | |
|---|---|---|---|---|---|---|
| | (I) | (II) | (III) | (IV) | (V) | (VI) |
| log(Unadjusted Distance) | -0.155** | | -0.184** | | -0.055** | |
| | [-2.34] | | [-2.55] | | [-2.09] | |
| log(TF Distance) | | -0.142 | | -0.359*** | | -0.044* |
| | | [-1.57] | | [-3.08] | | [-1.83] |
| Merger Completed | 0.021 | 0.022 | 0.053** | 0.053** | 0.007 | 0.008 |
| | [0.93] | [0.93] | [2.08] | [2.12] | [0.54] | [0.56] |
| Diversify | 0.040* | 0.020 | 0.023 | 0.005 | 0.008 | 0.001 |
| | [1.81] | [1.11] | [0.97] | [0.28] | [1.15] | [0.19] |
| Hostile | 0.073 | 0.076* | 0.133*** | 0.139*** | 0.028** | 0.029** |
| | [1.66] | [1.80] | [3.37] | [3.66] | [2.06] | [2.21] |
| High Tech | 0.005 | 0.007 | -0.023 | -0.023 | -0.011* | -0.010* |
| | [0.20] | [0.27] | [-1.00] | [-1.00] | [-1.91] | [-1.72] |
| Tender Offer | 0.103*** | 0.103*** | 0.157*** | 0.156*** | 0.024*** | 0.025*** |
| | [3.26] | [3.28] | [4.51] | [4.56] | [3.09] | [3.18] |
| Stock Deal | 0.071 | 0.075 | 0.096 | 0.103 | 0.061* | 0.062* |
| | [1.48] | [1.61] | [1.61] | [1.69] | [1.91] | [1.95] |
| Relative Deal Size | 0.000*** | 0.000*** | 0.000*** | 0.000*** | 0.000 | 0.000 |
| | [5.37] | [5.35] | [2.78] | [2.80] | [0.28] | [0.36] |
| Firm Size (Acquiror) | 0.018*** | 0.018*** | 0.029*** | 0.028*** | -0.013*** | -0.012*** |
| | [3.87] | [3.92] | [4.43] | [4.54] | [-6.32] | [-6.40] |
| Tobin's Q (Acquiror) | -0.001 | -0.001 | -0.000 | -0.000 | -0.000 | -0.000 |
| | [-0.28] | [-0.24] | [-0.05] | [-0.06] | [-0.08] | [-0.06] |
| Book Leverage (Acquiror) | -0.062* | -0.067* | 0.006 | 0.001 | 0.002 | 0.001 |
| | [-1.78] | [-1.94] | [0.14] | [0.01] | [0.11] | [0.06] |
| Cash Flow-to Asset (Acquiror) | 0.020 | 0.022 | -0.057* | -0.060* | -0.045** | -0.044** |
| | [1.11] | [1.33] | [-1.86] | [-1.93] | [-2.30] | [-2.12] |
| Firm Size (Target) | -0.033*** | -0.033*** | -0.054*** | -0.053*** | 0.009*** | 0.009*** |
| | [-4.98] | [-4.95] | [-5.31] | [-5.23] | [3.16] | [3.19] |
| Tobin's Q (Target) | -0.000 | -0.000 | -0.000 | -0.000 | 0.000 | 0.000 |
| | [-0.97] | [-0.89] | [-0.16] | [-0.24] | [0.16] | [0.08] |
| Book Leverage (Target) | 0.044 | 0.044 | 0.011 | 0.015 | -0.016 | -0.017 |
| | [0.70] | [0.71] | [0.19] | [0.24] | [-1.18] | [-1.21] |
| Cash Flow to Asset (Target) | -0.117 | -0.114 | 0.023 | -0.038 | 0.004 | -0.008 |
| | [-0.62] | [-0.58] | [0.04] | [-0.06] | [0.03] | [-0.07] |
| Intercept | 0.140** | 0.159** | 0.241*** | 0.293*** | 0.072*** | 0.077*** |
| | [2.47] | [2.66] | [3.21] | [3.79] | [2.75] | [2.89] |
| | | | | | | |
| Acquiror-Target Weighting | Equal | Equal | Equal | Equal | Value | Value |
| Year FE | Yes | Yes | Yes | Yes | Yes | Yes |
| Industry FE | Yes | Yes | Yes | Yes | Yes | Yes |
| Observations | 1,235 | 1,235 | 1,003 | 1,003 | 1,003 | 1,003 |
| R-squared | 0.126 | 0.124 | 0.196 | 0.199 | 0.141 | 0.138 |

### 4.5.2 Post-Merger Survival

Next, we investigate a relatively longer horizon effect of mergers. If two organizations are smoothly and efficiently integrated under one roof, the combined firm is more likely to survive following a M&A. To capture this medium to long term effect of M&As, we run probit regressions with acquiror survival one to two years post-M&A as our main dependent variable.

Control variables are similar to those employed in our deal-level analyses, Tables 4 and 9. Controlling for various deal characteristics and acquiror and target characteristics, we find that for a marginal increase in both of our functional distance measures, there is an 8.7% to 13.7% reduction in survival likelihood. These marginal effects are reported in Panel A of Table 10 below. Panel B reports the point estimates for all our explanatory variables.

### Table 10. Post-Merger Real Effects (SDC Platinum Deal Level: 1990-2021): Post-Merger Acquiror Survival

This table shows the relationship between the likelihood of post-merger acquiror survival and industry distance measures from 1990 to 2021. The dependent variable is an indicator for post-merger acquiror survival that is one if the acquiror exists after 1 year (Columns I, II, V, VI, IX, and X) and 2 years (Columns III, IV, VII, VIII, XI, and XII). The key independent variables are log(Unadjusted Distance) in Columns I, III, V, VII, IX, and XI and log(TF Distance) in Columns II, IV, VI, VIII, X, and XII. Columns (I) to (IV) are probit regressions for deals involving both public and private firms (acquiror and target). Columns (V) to (VIII) are probit regressions for deals involving publicly traded acquirors. Columns (IX) and (XII) are probit regressions for deals involving publicly traded acquirors with publicly traded targets. All regressions are controlled for Diversify, Hostile, High Tech, Tender Offer, Stock Deal, and Relative Deal Size. Additionally, Columns (V) to (VIII) control for acquirors' Firm Size, Tobin's Q, Book Leverage, and Cash Flow to Assets ratio and year and industry fixed effects; Columns (IX) to (XII) control for both acquirors' and targets' Firm Size, Tobin's Q, Book Leverage, and Cash Flow to Assets ratio as well as year and industry fixed effects. The t-statistics are shown in square brackets and standard errors are clustered at the year level. ***, **, and * denote significance at the 1%, 5%, and 10% level, respectively.

# Table 10 (continued)

Panel A. Marginal Effect

| Sample | All Deals | | | | Deals with Public Acquirors | | | | Deals with Public Acquirors-Public Target | | | |
|---|---|---|---|---|---|---|---|---|---|---|---|---|
| Forecast Horizon | t+1 | t+1 | t+2 | t+2 | t+1 | t+1 | t+2 | t+2 | t+1 | t+1 | t+2 | t+2 |
|  | (I) | (II) | (III) | (IV) | (V) | (VI) | (VII) | (VIII) | (IX) | (X) | (XI) | (XII) |
| log(Unadjusted Distance) | -0.087*** |  | -0.124*** |  | -0.060** |  | -0.090*** |  | 0.073 |  | -0.007 |  |
|  | [-3.36] |  | [-4.34] |  | [-2.10] |  | [-2.77] |  | [1.31] |  | [-0.10] |  |
| log(TF Distance) |  | -0.095*** |  | -0.137*** |  | -0.091** |  | -0.135** |  | 0.037 |  | -0.039 |
|  |  | [-2.85] |  | [-2.73] |  | [-2.41] |  | [-2.50] |  | [0.46] |  | [-0.39] |
| Deal Completed | 0.016* | 0.016* | 0.025** | 0.025** | 0.005 | 0.005 | 0.012 | 0.012 | 0.039*** | 0.039*** | 0.018 | 0.018 |
|  | [1.76] | [1.78] | [2.49] | [2.50] | [0.50] | [0.49] | [1.01] | [1.00] | [3.29] | [3.19] | [1.16] | [1.16] |
| Diversify | 0.008 | -0.003 | 0.027*** | 0.013 | 0.010 | 0.003 | 0.029*** | 0.019* | -0.002 | 0.006 | -0.000 | -0.000 |
|  | [1.11] | [-0.35] | [2.84] | [1.28] | [1.35] | [0.46] | [2.62] | [1.84] | [-0.18] | [0.85] | [-0.03] | [-0.02] |
| Hostile | 0.063* | 0.064* | 0.065* | 0.066* | 0.083** | 0.084** | 0.070* | 0.070* | 0.080** | 0.079** | 0.057 | 0.057 |
|  | [1.73] | [1.76] | [1.69] | [1.70] | [2.23] | [2.26] | [1.84] | [1.84] | [2.31] | [2.28] | [1.48] | [1.48] |
| High Tech | 0.005 | 0.007 | 0.021* | 0.025** | 0.000 | 0.001 | 0.013 | 0.015 | 0.013 | 0.011 | 0.028 | 0.028 |
|  | [0.49] | [0.74] | [1.70] | [2.04] | [0.00] | [0.06] | [0.99] | [1.11] | [1.03] | [0.92] | [1.25] | [1.24] |
| Tender Offer | 0.021* | 0.021* | 0.044*** | 0.045*** | 0.001 | 0.000 | 0.014 | 0.013 | -0.009 | -0.009 | 0.000 | -0.000 |
|  | [1.93] | [1.94] | [3.12] | [3.11] | [0.06] | [0.03] | [0.94] | [0.91] | [-0.71] | [-0.70] | [0.01] | [-0.01] |
| Stock Deal | -0.039 | -0.040 | 0.010 | 0.011 | -0.033 | -0.032 | -0.002 | 0.000 | 0.027 | 0.026 | 0.014 | 0.015 |
|  | [-1.50] | [-1.45] | [0.53] | [0.60] | [-1.51] | [-1.46] | [-0.07] | [0.01] | [0.79] | [0.76] | [0.31] | [0.32] |
| Relative Deal Size | 0.002 | 0.003 | -0.000 | -0.000 | 0.002 | 0.002 | -0.000 | -0.000 | 0.003 | 0.002 | -0.002 | -0.002 |
|  | [1.14] | [1.25] | [-0.18] | [-0.21] | [0.99] | [1.06] | [-0.17] | [-0.20] | [0.59] | [0.56] | [-0.41] | [-0.41] |
| Firm Size (Acquiror) |  |  |  |  | 0.013*** | 0.013*** | 0.021*** | 0.021*** | 0.006* | 0.005* | 0.020*** | 0.020*** |
|  |  |  |  |  | [6.18] | [6.22] | [9.59] | [9.62] | [1.86] | [1.82] | [4.66] | [4.68] |
| Tobin's Q (Acquiror) |  |  |  |  | -0.000** | -0.000** | 0.000 | 0.000 | 0.001 | 0.001 | 0.004* | 0.004* |
|  |  |  |  |  | [-2.20] | [-2.26] | [0.29] | [0.23] | [0.50] | [0.53] | [1.76] | [1.73] |
| Book Leverage (Acquiror) |  |  |  |  | 0.001 | 0.001 | -0.005*** | -0.005*** | 0.039 | 0.041 | 0.006 | 0.006 |
|  |  |  |  |  | [0.69] | [0.71] | [-3.83] | [-3.94] | [1.42] | [1.47] | [0.22] | [0.22] |
| Cash Flow to Asset (Acquiror) |  |  |  |  | 0.000 | 0.000 | -0.000*** | -0.000*** | 0.166*** | 0.165*** | 0.155*** | 0.155*** |
|  |  |  |  |  | [1.19] | [1.18] | [-3.13] | [-3.16] | [4.39] | [4.44] | [3.82] | [3.84] |
| Firm Size (Target) |  |  |  |  |  |  |  |  | -0.000 | -0.000 | -0.001 | -0.000 |
|  |  |  |  |  |  |  |  |  | [-0.09] | [-0.11] | [-0.14] | [-0.11] |
| Tobin's Q (Target) |  |  |  |  |  |  |  |  | 0.007*** | 0.008*** | 0.002 | 0.002 |
|  |  |  |  |  |  |  |  |  | [3.21] | [3.17] | [0.93] | [0.93] |
| Book Leverage (Target) |  |  |  |  |  |  |  |  | 0.038 | 0.038 | 0.078*** | 0.078*** |
|  |  |  |  |  |  |  |  |  | [1.50] | [1.50] | [2.87] | [2.90] |
| Cash Flow to Asset (Target) |  |  |  |  |  |  |  |  | -0.040 | -0.038 | -0.034 | -0.034 |
|  |  |  |  |  |  |  |  |  | [-0.67] | [-0.65] | [-0.58] | [-0.58] |
| Model | Probit | Probit | Probit | Probit | Probit | Probit | Probit | Probit | Probit | Probit | Probit | Probit |
| Public Acquiror | No | No | No | No | Yes | Yes | Yes | Yes | Yes | Yes | Yes | Yes |
| Public Target | No | No | No | No | No | No | No | No | Yes | Yes | Yes | Yes |
| Year FE | Yes | Yes | Yes | Yes | Yes | Yes | Yes | Yes | Yes | Yes | Yes | Yes |
| Industry FE | Yes | Yes | Yes | Yes | Yes | Yes | Yes | Yes | Yes | Yes | Yes | Yes |
| Pseudo R2 | 0.037 | 0.036 | 0.034 | 0.033 | 0.063 | 0.064 | 0.058 | 0.058 | 0.137 | 0.136 | 0.113 | 0.113 |
| Observations | 14,939 | 14,939 | 14,666 | 14,666 | 11,493 | 11,493 | 11,266 | 11,266 | 2,935 | 2,935 | 3,130 | 3,130 |

# Table 10 (continued)

Panel B. Coefficients

| Sample | All Deals | | | | Deals with Public Acquirors | | | | Deals with Public Acquirors-Public Target | | | |
|---|---|---|---|---|---|---|---|---|---|---|---|---|
| Forecast Horizon | t+1 | t+1 | t+2 | t+2 | t+1 | t+1 | t+2 | t+2 | t+1 | t+1 | t+2 | t+2 |
|  | (I) | (II) | (III) | (IV) | (V) | (VI) | (VII) | (VIII) | (IX) | (X) | (XI) | (XII) |
| log(Unadjusted Distance) | -0.646*** |  | -0.556*** |  | -0.410* |  | -0.394*** |  | 0.633 |  | -0.059 |  |
|  | [-3.34] |  | [-4.32] |  | [-1.92] |  | [-2.66] |  | [1.35] |  | [-0.17] |  |
| log(TF Distance) |  | -0.709*** |  | -0.622*** |  | -0.673** |  | -0.624** |  | 0.242 |  | -0.258 |
|  |  | [-2.85] |  | [-2.75] |  | [-2.35] |  | [-2.47] |  | [0.34] |  | [-0.47] |
| Deal Completed | 0.117* | 0.118* | 0.112** | 0.112** | 0.049 | 0.049 | 0.054 | 0.053 | 0.388*** | 0.380*** | 0.109 | 0.108 |
|  | [1.77] | [1.79] | [2.48] | [2.49] | [0.65] | [0.64] | [0.97] | [0.96] | [3.57] | [3.49] | [1.22] | [1.22] |
| Diversify | 0.056 | -0.021 | 0.120*** | 0.057 | 0.061 | 0.018 | 0.126*** | 0.085* | 0.044 | 0.116* | 0.034 | 0.033 |
|  | [1.09] | [-0.36] | [2.84] | [1.27] | [1.12] | [0.35] | [2.65] | [1.89] | [0.59] | [1.83] | [0.44] | [0.52] |
| Hostile | 0.465* | 0.472* | 0.296* | 0.297* | 0.639** | 0.645** | 0.335* | 0.334* | 1.010*** | 0.990*** | 0.390 | 0.392 |
|  | [1.73] | [1.76] | [1.70] | [1.71] | [2.29] | [2.32] | [1.89] | [1.88] | [3.03] | [2.91] | [1.64] | [1.64] |
| High Tech | 0.035 | 0.051 | 0.097* | 0.111** | 0.019 | 0.021 | 0.076 | 0.080 | 0.168 | 0.144 | 0.187 | 0.184 |
|  | [0.48] | [0.73] | [1.72] | [2.04] | [0.25] | [0.29] | [1.24] | [1.36] | [1.58] | [1.47] | [1.61] | [1.59] |
| Tender Offer | 0.154* | 0.156* | 0.191*** | 0.192*** | 0.015 | 0.013 | 0.059 | 0.057 | -0.070 | -0.072 | -0.004 | -0.005 |
|  | [1.93] | [1.93] | [3.04] | [3.04] | [0.17] | [0.15] | [0.90] | [0.87] | [-0.57] | [-0.58] | [-0.04] | [-0.06] |
| Stock Deal | -0.288 | -0.294 | 0.045 | 0.051 | -0.249 | -0.246 | -0.010 | -0.001 | 0.258 | 0.251 | 0.101 | 0.104 |
|  | [-1.49] | [-1.45] | [0.53] | [0.61] | [-1.53] | [-1.47] | [-0.09] | [-0.01] | [0.80] | [0.78] | [0.39] | [0.40] |
| Relative Deal Size | 0.017 | 0.019 | -0.000 | -0.000 | 0.017 | 0.018 | -0.000 | -0.000 | 0.027 | 0.025 | -0.010 | -0.010 |
|  | [1.14] | [1.25] | [-0.18] | [-0.21] | [1.04] | [1.10] | [-0.15] | [-0.18] | [0.67] | [0.65] | [-0.37] | [-0.38] |
| Firm Size (Acquiror) |  |  |  |  | 0.093*** | 0.095*** | 0.093*** | 0.095*** | 0.035 | 0.035 | 0.106*** | 0.106*** |
|  |  |  |  |  | [5.64] | [5.68] | [9.02] | [9.05] | [1.42] | [1.37] | [4.24] | [4.24] |
| Tobin's Q (Acquiror) |  |  |  |  | -0.000** | -0.000** | 0.000 | 0.000 | 0.009 | 0.010 | 0.025* | 0.025* |
|  |  |  |  |  | [-2.29] | [-2.35] | [0.28] | [0.22] | [0.50] | [0.53] | [1.71] | [1.68] |
| Book Leverage (Acquiror) |  |  |  |  | 0.004 | 0.004 | -0.022*** | -0.022*** | 0.365 | 0.381 | 0.002 | 0.002 |
|  |  |  |  |  | [0.66] | [0.68] | [-3.54] | [-3.65] | [1.37] | [1.41] | [0.01] | [0.01] |
| Cash Flow to Asset (Acquiror) |  |  |  |  | 0.000 | 0.000 | -0.000*** | -0.000*** | 1.600*** | 1.587*** | 0.895*** | 0.892*** |
|  |  |  |  |  | [1.19] | [1.18] | [-2.93] | [-2.96] | [4.30] | [4.33] | [3.68] | [3.69] |
| Firm Size (Target) |  |  |  |  |  |  |  |  | 0.024 | 0.024 | -0.002 | -0.001 |
|  |  |  |  |  |  |  |  |  | [0.98] | [0.95] | [-0.08] | [-0.05] |
| Tobin's Q (Target) |  |  |  |  |  |  |  |  | 0.068*** | 0.070*** | 0.012 | 0.012 |
|  |  |  |  |  |  |  |  |  | [3.19] | [3.14] | [0.82] | [0.82] |
| Book Leverage (Target) |  |  |  |  |  |  |  |  | 0.295 | 0.292 | 0.425*** | 0.428*** |
|  |  |  |  |  |  |  |  |  | [1.33] | [1.32] | [2.80] | [2.83] |
| Cash Flow to Asset (Target) |  |  |  |  |  |  |  |  | -0.508 | -0.485 | -0.225 | -0.224 |
|  |  |  |  |  |  |  |  |  | [-0.91] | [-0.89] | [-0.65] | [-0.64] |
| Intercept | 1.553*** | 1.702*** | 1.328*** | 1.461*** | 1.110*** | 1.259*** | 0.882*** | 1.021*** | 2.367*** | 2.332*** | 1.364*** | 1.424*** |
|  | [12.49] | [12.37] | [14.03] | [14.83] | [7.22] | [7.08] | [8.76] | [8.65] | [3.46] | [3.14] | [5.64] | [4.79] |
| Model | Probit | Probit | Probit | Probit | Probit | Probit | Probit | Probit | Probit | Probit | Probit | Probit |
| Public Acquiror | No | No | No | No | Yes | Yes | Yes | Yes | Yes | Yes | Yes | Yes |
| Public Target | No | No | No | No | No | No | No | No | Yes | Yes | Yes | Yes |
| Year FE | Yes | Yes | Yes | Yes | Yes | Yes | Yes | Yes | Yes | Yes | Yes | Yes |
| Industry FE | Yes | Yes | Yes | Yes | Yes | Yes | Yes | Yes | Yes | Yes | Yes | Yes |
| Pseudo R2 | 0.037 | 0.036 | 0.034 | 0.033 | 0.063 | 0.064 | 0.058 | 0.058 | 0.137 | 0.136 | 0.113 | 0.113 |
| Observations | 14,939 | 14,939 | 14,666 | 14,666 | 11,493 | 11,493 | 11,266 | 11,266 | 2,935 | 2,935 | 3,130 | 3,130 |

Overall, heterogeneous production decision processes between acquirors and targets are strongly correlated with poor valuation outcomes reflected in the short- and long-term analyses. The results indicate higher integration costs and lower synergy in mergers across heterogeneous industries.

## 5. Conclusion

We analyze the impact of production process heterogeneity between acquiror and target in the M&A market. Using deep learning techniques, we estimate each firm's production decision process at the industry level. We use various firm characteristics as production inputs and map them to production outcomes non-parametrically and non-linearly. Machine learning determines on these non-linear transformations of the inputs to the final production outcome. Once we estimate the weights that map the production inputs to the output, we compute the functional distances between acquirors' and targets' industries. Using the Lagrange multiplier-like intuition, we compute the ratio of the mean squared errors in the constrained production function estimation to the mean squared errors of the unconstrained case. This mean squared errors ratio captures how difficult it is for an acquiror to integrate the target's production process under one roof. When we construct these functional distances, we apply not only a native deep learning algorithm but also a more advanced Transfer Learning algorithm that conducts "layer extraction" from the trained industry (i.e., acquiror's industry) to the test industry (i.e., target's industry). Using the two functional distances estimated by the two deep neural network approaches, we examine our key hypotheses regarding their effects on merger activities and outcomes.

We test whether an acquiror and target from industries with large differences in production decision processes are less likely to undergo business integration via M&As. We further examine whether those deals are less likely to be successfully completed and and if they exhibit lower stock market returns when they are announced to investors. We also investigate whether firms with greater functional distances are less likely to survive in the one- and two-year horizon following merger completion. When we conduct these tests, we also conjecture that Transfer Learning-based distance is more likely to capture salient differences in production process heterogeneity across industries relative to the native deep learning-based distance.

Using nearly 30 years of comprehensive M&A data for U.S. firms compiled from the SDC Platinum database since 1990, we find empirical results that are largely consistent with our main predictions; M&As between firms from distinct industries with heterogeneous production factors and decision-making rules, culture, and organizational structures tend to be unsuccessful. There are significantly lower numbers of M&As between distinct industries in their production functions and the deals are much less likely to be completed. Stock market reactions to the announcement of the deals are 3% to 5% worse, and these firms have an 8.7% to 13.7% lower survival likelihood one and two years following the merger. All these results indicate lower merger synergy between firms from heterogeneous industries in terms of their production functional profiles.

In our robustness checks, we show results for an alternative production outcome; ROA, replacing Tobin's Q, generates results that are virtually unchanged from our earlier findings. Put together, for the first time in the literature we quantify production process heterogeneity across industries and show that it is an important determinant of merger activities and the short- and long-term performance outcome of the mergers.

Our deep learning algorithms attempt to capture internal production decision processes that are latent by nature. Production factors and weights that we implicitly identify via deep neural networks here could be closely related to both the underlying production technology and other intangible capital that the merging organizations have, such as culture, organizational capital, and decision hierarchies. For example, some companies heavily rely on top managers' decisions in the last round of their production decision tree, while others rely more heavily on the lower layers of the tree: they are more open to lower ranked employees' entrepreneurial ideas and their business executions. Our production process estimates could explain many such elements of corporate financial decision making and outcomes beyond M&As. They might also be able to explain managerial decision-making preferences and patterns of CEO turnovers, i.e., whether new CEOs are internally promoted or from outside the company, and which type of CEO would be more suitable for a given organization. We hope to return to these questions in subsequent research.